\pdfoutput=1
\documentclass[a4paper,fleqn,usenatbib]{mnras}

\usepackage{newtxtext,newtxmath}
\usepackage[T1]{fontenc}
\usepackage{ae,aecompl}

\usepackage{graphicx}	% Including figure files
\usepackage{amsmath}	% Advanced maths commands
\usepackage{amssymb}	% Extra maths symbols

\title[MQs as sources of IGM heating and ionisation]{Microquasars as heating sources of the intergalactic medium during reionisation of the Universe}

\author[V. M. Douna et al.]{
Vanesa M. Douna,$^{1,2}$\thanks{E-mail: vdouna@iafe.uba.ar}
Leonardo J. Pellizza,$^{3}$
Philippe Laurent,$^{4,5}$
I. F\'elix Mirabel.$^{2,4}$
\\
$^{1}$ Universidad de Buenos Aires, Facultad de Ciencias Exactas y Naturales, Departamento de F\'isica. Buenos Aires, Argentina.\\
$^{2}$ CONICET-Universidad de Buenos Aires, Instituto de Astronom\'ia y F\'isica del Espacio (IAFE). Buenos Aires, Argentina.\\
$^{3}$ Instituto Argentino de Radioastronom\'ia, (CICPBA -- CONICET). Villa Elisa, Argentina.\\
$^{4}$ CEA-Saclay, IRFU/DSM/Service d' Astrophysique, 91191 Gif-sur-Yvette, France.\\
$^{5}$ Astroparticules et Cosmologie (APC), CEA/IRFU, 75205, Paris Cedex 13, France.\\
}

\date{Accepted XXX. Received YYY; in original form ZZZ}

\pubyear{2017}

\begin{document}
\label{firstpage}
\pagerange{\pageref{firstpage}--\pageref{lastpage}}
\maketitle

\begin{abstract}
The effect of the primeval sources of radiation and particles on the thermodynamical state of the intergalactic medium during the Epoch of Reionisation is still unclear. In this work, we explore the contribution of electrons accelerated in the jets of high-redshift microquasars to heating and ionising the intergalactic medium. We develop Monte Carlo simulations of the propagation and energy deposition of these electrons as they travel away from their sources. We find that microquasars contribute significantly to heating the intergalactic medium and are effective ionisers only near the galaxies. Their effect on heating is of the same order of magnitude than that of CRs from SNe.

\end{abstract}


%%%%%%%%%%%%%%%%%%%%%%%%%%%%%%%%%%%%%%%%%%%%%%%%%%

%%%%%%%%%%%%%%%%% BODY OF PAPER %%%%%%%%%%%%%%%%%%

\section{Introduction}

Reioisation is one of the major phase transitions of the Universe. At $z \approx 1100$ ($0.37\,{\rm Myr}$ after the Big Bang) the plasma in the Universe became neutral and decoupled from radiation, releasing the Cosmic Microwave Background (CMB). However, it has been shown that $\sim 1\,{\rm Gyr}$ after the Big Bang, the intergalactic medium (IGM) was warm and ionised again, evidencing a phase transition named `Epoch of Reionisation' (EoR) that is believed to have occured after the birth of the first stars and galaxies. 

Many efforts have been made in order to characterise the effect of the first structures in the process of hydrogen and helium ionisation, as well as in the thermal history of the IGM \citep[and references therein]{barkanaloeb2001,ciardiferrara2005,pritchardfurlanetto2007,fialkovbarkana2014,loeb2010,zaroubi2013,mesinger2016}.  However, the sources of this phase transition are still under debate, as well as the detailed temporal and spatial structure of the process. 

The development of the observational skills during the last decade have placed constraints to the timeline of the EoR. Observations of the Ly$\alpha$ absorption lines in the spectra of high redshift quasars suggested that reionisation was completed at redshift $z \approx 5-6$ \citep{fan2006,becker2015}. On the other hand, CMB polarization measurements published by the \citet{Planck2016} are consistent with an average redshift of $z \approx 7.8-8.8$ (depending on the model) and a maximal duration of $\Delta z<2.8$, reducing the preexisting tension between the CMB experiments and the measurements from high-redshift astrophysical sources. 

The mostly accepted agents of reionisation are massive stars \citep[e.g.,][]{robertson2010}, which emit UV radiation capable of ionising \ion{H}{i}. However, evidence suggests that the ionising radiation emitted by massive stars in high-redshift galaxies is not enough to keep the IGM ionised. To account for the missing UV radiation, either the escape fraction of ionising photons should have been higher than the observed rate \citep{heckman2001,mitra2013,ferraraloeb2013,izotov2016}, or a population of galaxies with luminosities under the detection limit is required \citep{lehnert2010, trenti2010, alvarez2012, wise2014}. It is believed that also AGNs have contributed to the process of reionisation, although their relative effect with respect to other sources at different redshifts is still being debated \citep[e.g.,][]{fan2001, cowie2009, madauhaardt2015}. The lack of agreement triggered a search for new reionisation sources.

Several authors have proposed that X-ray binaries (XRBs) could contribute to IGM heating and reionisation \citep{mirabel2011,fragos2013,xu2014,jeon2014,sazonov2017}. Despite the scarcity and short lifetimes of these sources, their luminosity, together with the large mean free path and ionising power of X-ray photons in the IGM, make XRBs attractive candidates. However, recent works suggest that the contribution of the X-rays emitted by these systems to heating and ionisation of the IGM during the EoR, is marginal at most \citep{madaufragos2017}.

Cosmic rays (CRs) have also been proposed as sources of heating and ionisation at high redshift \citep{nath1993}. It has been suggested that CRs accelerated in supernova explosions (SNe) could contribute to heating the IGM \citep[e.g.][]{samui2005}. In particular, \citet{sazonovsunyaev2015}, have shown that low-energy cosmic rays may have taken a significant fraction of the kinetic energy of the SNe originated from the first generation (Pop III) of stars, and that they may have been responsible of heating the IGM at high redshift before other sources. However, Pop III stars could have been so massive that they could also have collapsed without a SNe \citep{mirabelrodrigues2003}. \citet{leite2017} have analysed the role of low-energy CRs accelerated in SNe from Pop II stars, showing that their contribution to ionisation would have been negligible, although they may have been effective heating sources at $z \sim 10$.

A subset of XRBs, known as microquasars \citep[MQs,][]{mirabelrodriguez1999}, exhibit powerful relativistic jets. Steady jets in MQs are emitted in the low-hard state of the spectrum and are mildly relativistic, while discrete outbursts are associated to fast ejections with ultrarelativistic velocities. The kinetic luminosity of jets strongly varies for different sources, from $10^{35}$ to $10^{40}\,{\rm erg\,s}^{-1}$ \citep{fabrika2004,gallo2005,pakull2010}. MQ jets can transport energy far away from the XRB \citep[several parsecs, or even more,][]{pakull2010}, where they interact with the ambient medium \citep{zealey1980,mirabel1992,marti2005}. This interaction generates a complex shock structure in the termination of the jet, where particles are accelerated generating a non-thermal population of particles and radiation, which may escape to the ISM as CRs \citep[and references therein]{heinzsunyaev2002,bordas2009,zhangfeng2011}. The composition of the jets is still being debated \citep[e.g.,][]{Romero14}. They could be leptonic (comprising only an electron-positron plasma and photons), or lepto-hadronic (comprising also baryons); therefore the composition of the CR component produced by MQs is also uncertain. The radio lobes observed in some MQs constitute strong evidence of the presence of at least accelerated electrons in the jet/ISM interface.

CRs injected by the jets of MQs into the ISM may have contributed to reionisation in a non-negligible way. Their effect may have been enhanced in the early Universe for several reasons. As pointed out by \citet{mirabel2011}, the total luminosity of XRBs (and therefore MQs) per unit of star formation rate increases at high redshift due to effects related to the metallicity of their parent stellar populations \citep{basuzych2013a,fragos2013,basuzych2013b}. In this sense, there is strong evidence that supports that XRBs are more numerous in low-metallicity environments \citep{mapelli2010,kaaret2011,brorby2014,douna2015,brorby2016}. Stars in low metallicity environments are also believed to give birth to more massive black holes \citep[e.g.,][]{belczynski2010} and favour an enhanced formation rate of XRBs \citep[e.g.,]{dray2006}. Moreover, the mass transfer during the XRB phase should have been mostly due to Roche-lobe overflow accretion \citep[e.g.,][]{linden2010}, which results in higher accretion rates and higher luminosities. Pop III stars are believed to be extremely metal poor and to have almost no winds. Although the formation rate and features of Pop III binary systems are uncertain, theoretical findings suggest that the population of XRBs formed from Pop III stars should have been more luminous than the actual ones \citep[e.g.,][]{ryu2016}. Consequently, jets of MQs in the first XRBs might have been more powerful than in the actual Universe.

The contribution of high-energy CRs accelerated in MQ jets to ionisation in the early Universe has been investigated by \citet{tueros2014}. Their results imply, however, that this contribution would have been at most of the same order of magnitude than that of the X-rays emitted by the same sources. The best case corresponds, according to these authors, to a fully leptonic jet accelerating only electrons. 

\citet{tueros2014} explored the ionisation of large volumes of the Universe.  They compute ionisation and heat locally as proportional to the energy lost by high-energy particles through their interaction with the medium. As MQs are scarce and short-lived sources, their effects on the short (i.e., kpc--Mpc) scale, as well as the inhomogeneity of the ionisation they produce, deserve exploration. At these scales, the energy deposition of high-energy particles, used to ionise the medium, can not be considered local. These authors also assume that MQ CRs are injected directly into the IGM. Jets of hundreds of parsecs reaching the IGM are probably an extreme case. Hence, it would be important to investigate the problem in the case that MQ jets inject CRs within primeval galaxies, and assess the electronic escape fraction in this case.

Based on the aforementioned arguments, in this paper we explore the ionisation and heating produced by MQ CRs in the early-Universe IGM, at short (kpc--Mpc) scales. As a first step, we have chosen to investigate the effect of CR electrons only, as these particles are expected to make the largest contribution \citep{tueros2014}. To this aim, we develop and apply Monte Carlo simulations of the propagation and energy deposition of electrons through the IGM, spanning fifteen orders of magnitude in energy, while treating relevant physical processes in a detailed and self-consistent way.

The organisation of the paper is as follows: Sect.~\ref{sect:sim} describes the physical processes relevant to the propagation of electrons, and the way they are implemented in our simulations, whereas Sect.~\ref{sect:lowe} discusses the results concerning those electrons directly responsible for ionisation and heating. Sect.~\ref{sect:galaxy} and \ref{sect:igm} show the results focusing respectively on the particles that manage to leave the galaxy into the IGM, and the ionisation and heating of the latter. Finally, Sect.~\ref{conclusions} discusses our results and shows our conclusions.

\section{Electron propagation simulations}
\label{sect:sim}

\subsection{General scenario}

Our aim is to describe the propagation of electrons through the ISM of a primordial galaxy, their escape into the IGM, and their contribution to ionising and heating the latter. We assume that electrons are produced in the jets of MQs and injected into the ISM, as discussed by \citet{heinzsunyaev2002}. After escaping from the source, electrons interact mainly with the ISM/IGM and any photon or magnetic field pervading the space through which they travel.

Both the ISM and IGM are assumed to be cold, homogeneous, partially ionised plasmas of primordial composition, $X=0.752$, $Y=0.248$, and $Z=0$ according to WMAP \citep{spergel2007}. As a first approach, we fix the working redshift at $z = 10$. This is thought to be the typical epoch at which first galaxies formed, and presumably that at which the main contribution to reionisation occurs \citep{Bromm09}. Previous evidence suggests that X ray radiation emitted by X-ray binaries (which appear in the Universe relatively late) might dominate over other sources and could affect the thermal balance only around that redshift \citep[e.g.][]{xu2014,madaufragos2017}. We adopt a typical value of $n_{\rm ISM} = 1\,{\rm cm}^{-3}$ for the ISM density, and the mean baryonic density of the Universe at redshift $z = 10$ for the IGM density ($n_{\rm IGM} = 2.4 \times 10^{-4}\,{\rm cm}^{-3}$). We leave the ionisation fraction of each medium, $f_{\rm ion} = n(\ion{H}{II})/n(\mathrm{H}) = n(\ion{He}{II})/n({\rm He})$, as a free parameter of our scenario. Here $n$ stands for the particle density of each component, and we assume that all ionised He is in the form of \ion{He}{ii}.

Photon fields may be produced by different sources. Large-scale fields (the CMB and the Extragalactic Backgroung Light -EBL-) seem to be {\em a priori} the most important for our work, as electrons interact with them all along their path. We model the CMB as a homogeneous and isotropic black body at a temperature $T = T_0 (1 + z)$, where $T_0 = 2.735\,{\rm K}$ is the present-day CMB temperature \citep{Fixsen09}. We do not model the EBL, because its energy density and spectrum at high redshift are highly uncertain \citep[e.g.,][and references therein]{Gilmore12}; as we will discuss in Sect.~\ref{conclusions}, this turns out to be a conservative choice. For the same reason, we disregard also photon fields from galactic sources (including those emitted by the same MQs that produce the electrons).

Magnetic fields may play an important role in the propagation of electrons, depending mainly on their intensity. The effects of magnetic fields are the cooling of electrons through synchrotron radiation, and the isotropisation of the electron directions of motion, acting as an effective diffusion mechanism. The investigation of the properties of galactic and cosmological magnetic fields is a relatively recent, but rapidly developing research area. However, there are still many uncertainties in the knowledge of these fields in the early Universe \citep[e.g.,][and references therein]{Durrer2013}. Therefore, we will treat magnetic fields separately in a companion paper.

As discussed above, in the present work we will restrict ourselves to the interactions of electrons with the CMB, the ISM, and the IGM. Electrons can excite or ionise neutral atoms in these plasmas, scatter off free electrons, or emit Bremsstrahlung radiation in the field of both atoms and ions. They can also interact with photons of the CMB through inverse Compton (IC) scattering. High-energy photons produced in these processes can further boost free ISM/IGM electrons through direct Compton scattering. Taking into account this set of interactions, we can model the ionisation produced and heat deposited in the IGM by the energy cascades arising from primary electrons. Part of the energy is carried away by photons, which can further photoionise the IGM plasma.  However, as a first step, we disregard it in order to save computational time. In any case, this is another conservative choice, as the inclusion of photoionisation would increase the ionising power of the sources.

The collisional ionisation cross section peaks at electron kinetic energies $E_{\rm k}$ of hundreds of eV, whereas the original electrons emitted by the sources can be as energetic as hundreds of TeV. This means that we must follow the evolution of particle energies through about twelve orders of magnitude, which requires a lot of computational time. However, the problem admits a natural separation that we exploit to avoid wasting computational time: below $E_{\rm k} \sim 10\,{\rm keV}$ ionisation, excitation, and elastic scattering processes dominate. For higher kinetic energies the main processes are Compton scattering and Bremsstrahlung. Therefore, we separate the simulation in low- and high-energy regimes, whose boundary we define as $E_{\rm k} = 10\,{\rm keV}$.

\subsection{Simulations in the low-energy regime}
\label{subsect:sim-LE}

For the low-energy regime, where ionisation plays a key role, we developed a Monte-Carlo code called {\em JET}. This code computes the propagation of a set of electrons with kinetic energies in the range $1\,{\rm eV}-10\,{\rm keV}$ in a partially ionised medium composed of H and He, by sampling their individual interactions as they move. {\em JET} incorporates a detailed treatment of H/He collisional ionisation and excitation, and recombinations. Heat deposition through $e^-e^-$ elastic scattering is computed as a continuous energy loss for each electron. Free-free interactions of electrons with either ionised or neutral atoms are not included as they are assumed negligible in comparison with other relevant processes \citep{valdes2010,valdesferrara2008}. 

The collisional ionisation cross sections for H and He are taken from \citet{shah1987,shah1988}. The energies of the secondary electrons generated from the ionising collisions are sampled from the distributions presented by \citet{opal1971}. The collisional excitation cross sections for H and He are from \citet{stone2002}. In each excitation process a photon is generated, and its energy is summed to a counter to keep track of the total energy lost to this channel, as the generated photon does not have enough energy to further contribute to ionisation.

The elastic collisions with free thermal electrons are simulated by means of a stopping power formalism \citep{spitzerscott1969,habing1971,shull1979}. The energy lost due to collisions with thermal electrons is transformed into heat. As a simplification, when the electron energy is lower than 10.2 eV, it is assumed that it will thermalise locally due to elastic $e^-e^-$ collisions, and its energy is therefore converted into heat. The radiative energy loss due to Bremsstrahlung is treated in the same way, following \citet{seltzer1985}.

The recombination mean free path is calculated using the cross sections of the free-bound transitions ($\sigma_{\rm fb}$) in a Coulomb potential, obtained from the photoionisation cross sections ($\sigma_{\rm bf}$) as in \citet{rybickilightman1979}. The resulting cross sections are

\begin{equation}
\sigma_{\rm fb}^n  \approx  1.05 \times 10^{-22} g_{\rm bf} \frac{\omega_n}{\omega_0} \frac{1}{n} \frac{I_n^2}{E E_e} \rm{cm^2},
  \label{eq:sigmafb}
\end{equation}

\noindent
where the subindex $n$ refers to the level of the electron in the recombined atom, $I_n$ is the ionisation energy corresponding to level $n$, $E$ is the energy of the outcoming photon and $E_e$ is the energy of the incoming electron. $\omega_n$ and $\omega_0$ are the statistical weights corresponding to the $n$ level of the recombined atom and the fundamental level of the recombining ion, respectively. The Gaunt factors $g_{\rm bf}$ were extracted from \cite{karzas1961}.

\subsection{Simulations in the high-energy regime}
\label{subsect:sim-HE}

In the high-energy regime, we have adapted for our purpose the code {\em UTOPIA} \citep[Understanding Transport of Particles In Astrophysics,][]{Pellizza10}. {\em UTOPIA} is esentially a Monte Carlo code for computing relativistic electromagnetic cascades produced by photons or electrons via IC and pair creation ($\gamma \gamma \to e^-e^+$) interactions. {\em UTOPIA} features a novel scheme, based on Markov-chain Monte Carlo (MCMC) integration and sampling, which is used to compute mean free paths and sample interaction products without requiring approximations of the corresponding cross sections. This implies that {\em UTOPIA} can be used to treat interactions in arbitrary fields, even those with no symmetries. On the other hand, it makes an efficient treatment of cascades in inhomogeneous fields. Our modification allows {\em UTOPIA} to treat the motion of electrons in the non-relativistic regime, namely at energies down to $E_{\rm k} = 3\,{\rm keV}$, and adds a detailed treatment of Bremsstrahlung, ionisation, and $e^-e^-$ elastic scattering.

IC and pair creation cross sections have been taken from \citet{Blumenthal70} and \citet{Gould67}, respectively. For the present work we have added the treatment of ionisation, Bremsstrahlung and elastic scattering. For the first process, we use the cross section of \citet{Kim00}, whereas the last two are computed as continuous energy losses, integrating their stopping powers instead of sampling individual interactions from the cross sections. The corresponding formulae have been taken from \citet{spitzerscott1969}, \citet{habing1971}, and \citet{shull1979} for elastic scattering, and from \citet{rybickilightman1979} for Bremsstrahlung.

We point out that the changes in the rates of the different processes due to the expansion of the Universe were disregarded, and that we have not taken into account the cosmological redshifting of the CMB photons for the calculation of the IC losses. In our simulations, energetic electrons travel at most 1~Mpc. This distance is rapidly covered, and the change in redshift corresponding to the time lag is small ($\Delta z \sim 0.05$ for a 1~MeV electron). The associated variations in the CMB temperature ($\Delta T / T \sim 0.005$) are then negligible. For this reason, the energy losses by IC stay almost unaffected. A detailed treatment of the redshifting \citep[e.g.][]{slatyer2010}, would be required only for large-scale simulations, or those covering large time spans.

\begin{figure}
\includegraphics[width=\columnwidth]{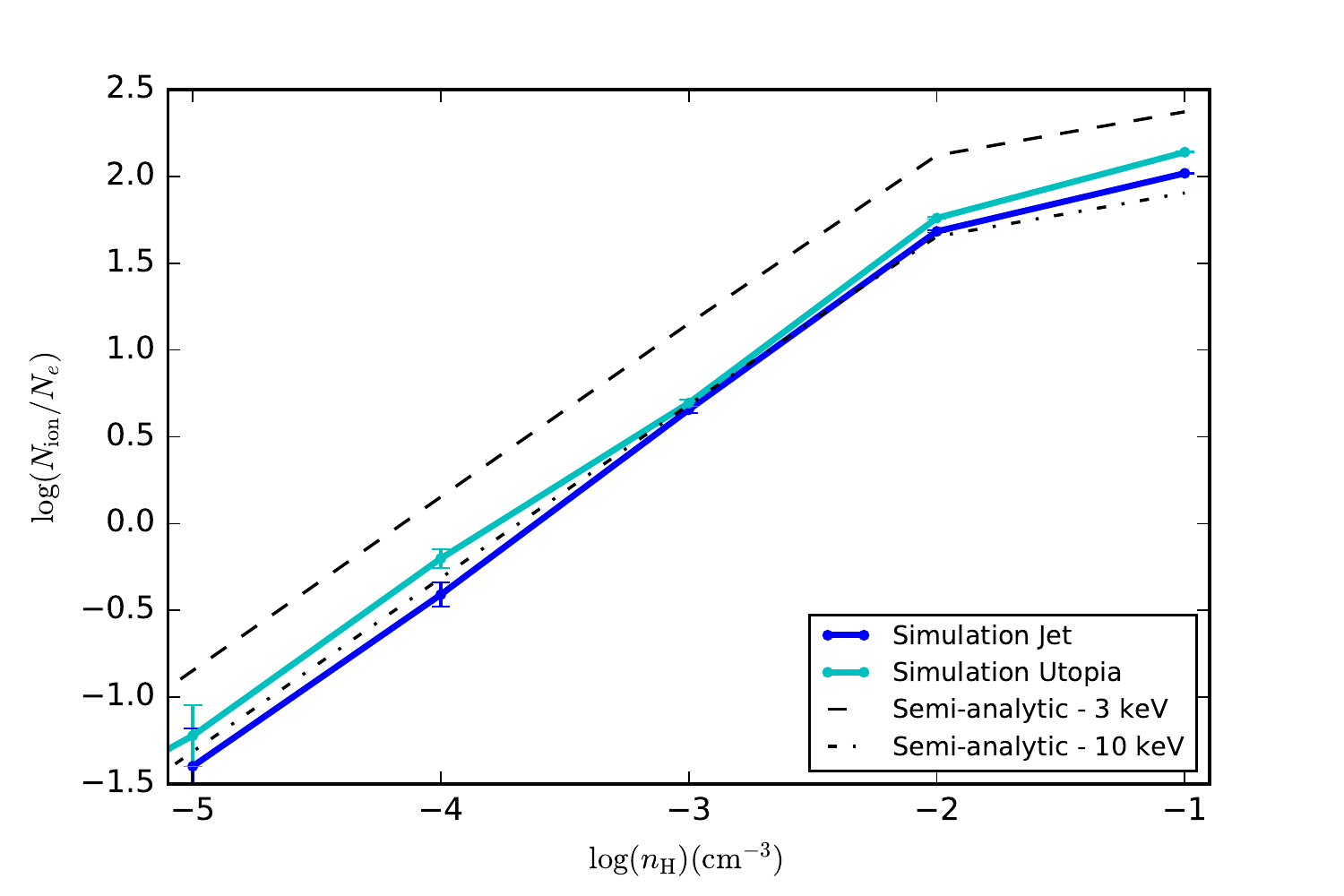}
\includegraphics[width=\columnwidth]{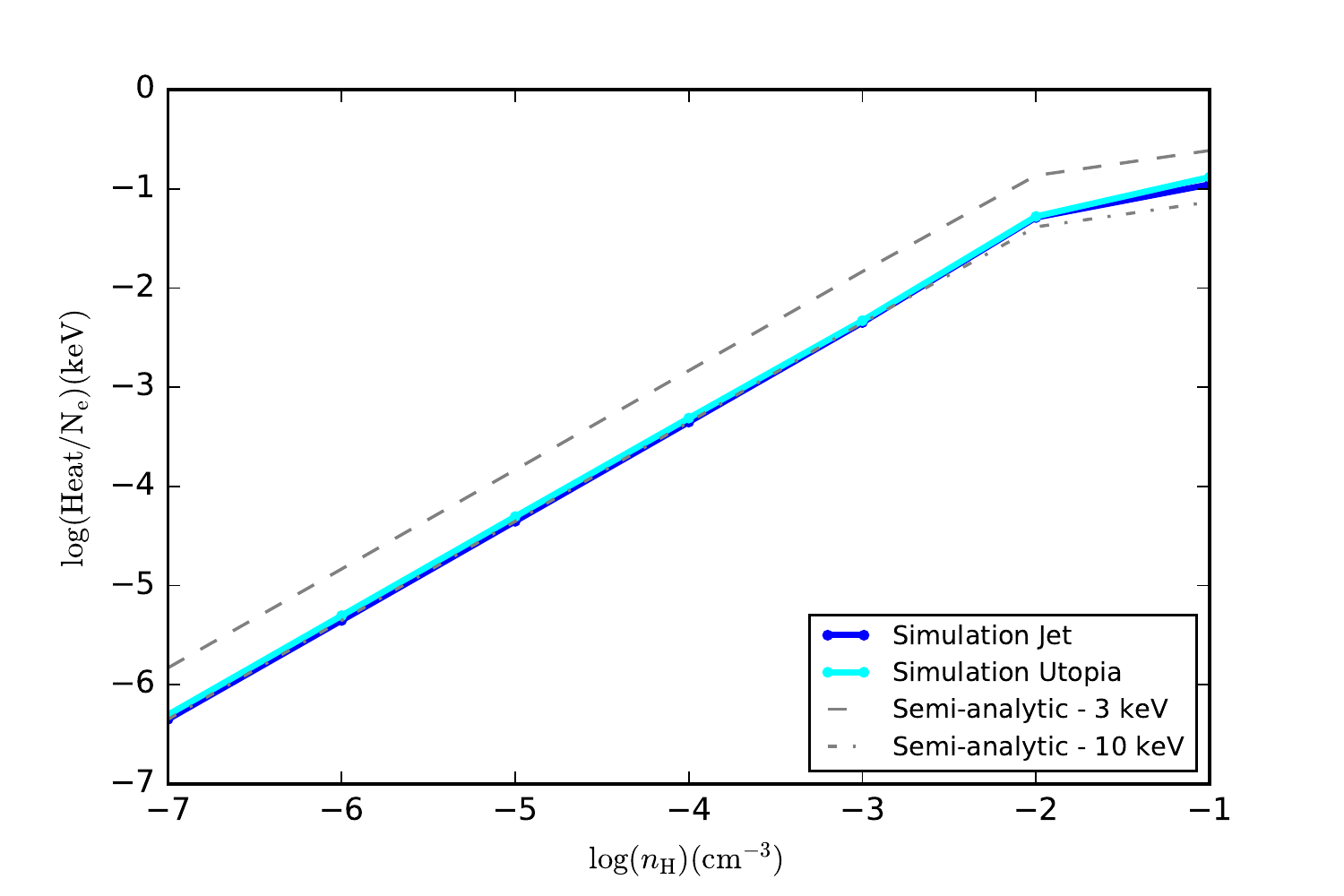}
\caption{Simulated number of ionisations (upper panel) and heat (lower panel)
produced within $1\,{\rm kpc}$ of the source, per primary electron of
$9.9\,{\rm keV}$, as a function of the density of the medium. The cyan and blue solid lines represent the {\em UTOPIA} and {\em JET} results,
respectively. Error bars are Poissonian uncertainties. The black
dash-dotted and dashed lines indicate the analytically predicted values
assuming that the particles travel at constant energies of $10$ and
$3\,{\rm keV}$, respectively (the limits of the overlapping energy range for
the two codes). {\em UTOPIA} and {\em JET} simulations are clearly consistent
with each other, and with analytical predictions.}
\label{fig:jet-utopia:Comparison}
\end{figure}

To check the consistency between {\em JET} and {\em UTOPIA}, we have run a set of simple simulations in which electrons of $9.9\,{\rm keV}$ travel outwards from the center of a $1\,{\rm kpc}$-radius sphere, through a homogeneous medium with a fixed ionisation fraction $f_{\rm ion} = 0.01$. The density of the medium has been varied from $10^{-7}$ to $10^{-1}\,{\rm cm}^{-3}$. In this computation we have used only the common physical processes, and followed the electrons until either their energy decreases below $3\,{\rm keV}$, or they reach the surface of the sphere. In this way, electrons are always in the energy range in which both codes overlap. As an example of the relative performance of the codes, Fig.~\ref{fig:jet-utopia:Comparison} shows a comparison of the number of ionisations and heat deposited in the medium as a function of its density. Analytical predictions of the same quantities, computed assuming that the particles travel at constant energies of $10$ and $3\,{\rm keV}$, are also shown. The consistency between both codes and the analytical predictions is clearly seen. As expected, the simulated curves agree with the $10\,{\rm keV}$ analytical one at low densities, because particles escape from the simulated sphere without losing much energy. At the highest densities probed, the mean distance travelled by electrons becomes lower than the size of the sphere, which explains the break in the curves. They also lose a significant amount of energy, making the simulated curves to move towards the $3\,{\rm keV}$ analytical prediction.

\section{Ionisation range of low-energy electrons}
\label{sect:lowe}

Most previous works \citep[e.g.,][]{shull1979,shull1985,valdesferrara2008,slatyer2010,furlanettostoever2010,valdes2010,evoli2012} concentrate on the global energetics of IGM ionisation. In order to assess the ionising power of individual sources such as MQs, we discuss also the the spatial distribution of the energy deposition in the medium. As stated in the previous section, the ionisation cross section peaks in the low-energy regime of our simulations, therefore we begin discussing the effect of the electrons in this regime ($E_{\rm k} < 10\,{\rm keV}$). All the results of this section have been obtained with the {\em JET} code alone. To this aim, we have run a set of simulations of a source of 1000 electrons with a primary kinetic energy $E_{\rm k}$, which travel outwards from a source at the center of a homogeneous sphere of radius $R$, particle density $n$ and ionisation fraction $f_{\rm ion}$. We have explored different values of $E_{\rm k} \in [0.3, 10]\,{\rm keV}$, $f_{\rm ion} \in [10^{-4},10^{-1}]$, and the density $n$ in a sufficiently large range to include both the typical values expected for the ISM and the IGM at $z = 10$. Electrons are followed until they lose their energy or reach the surface of the sphere.

\begin{figure}
\includegraphics[width=0.93\columnwidth]{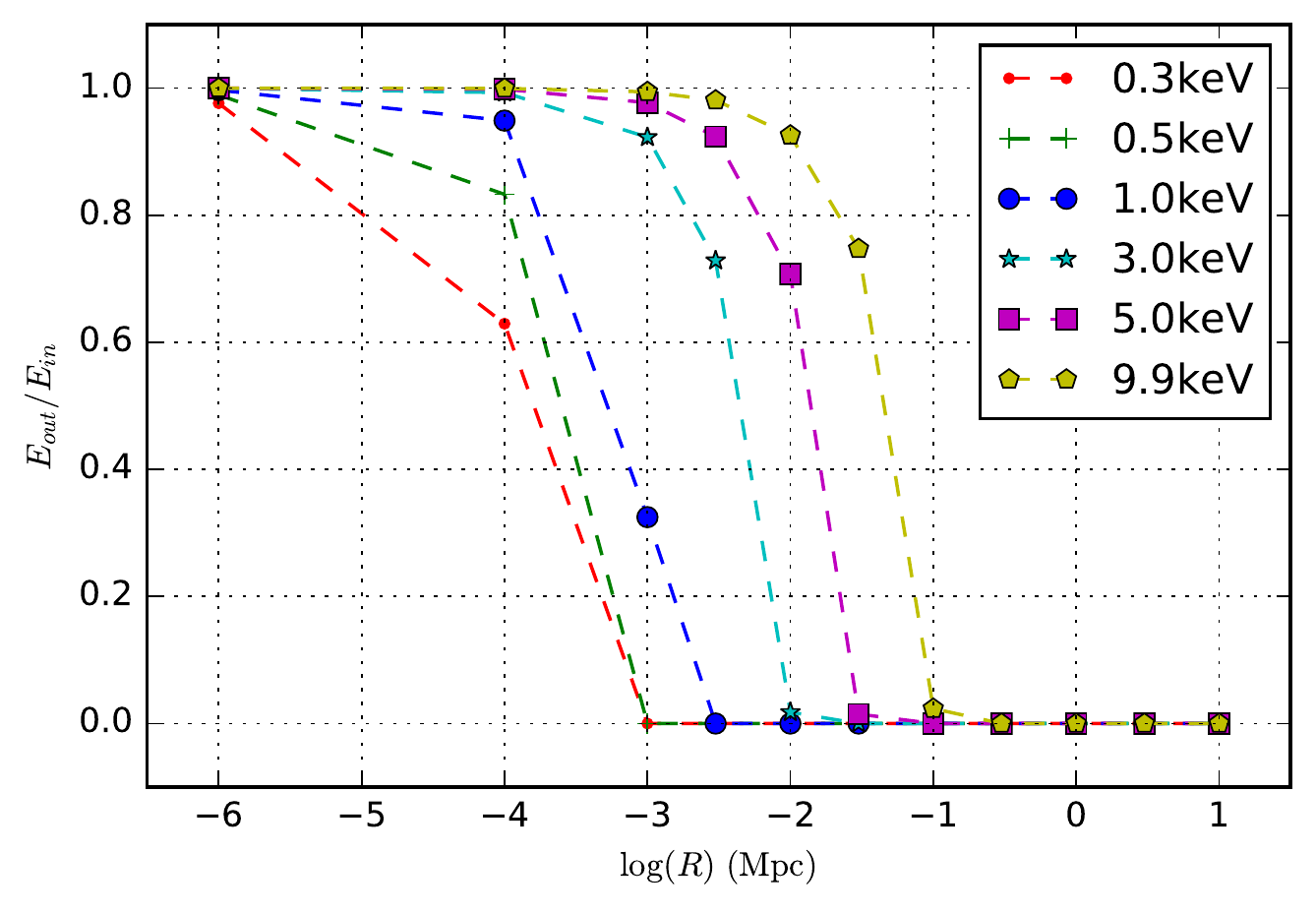}
\includegraphics[width=\columnwidth]{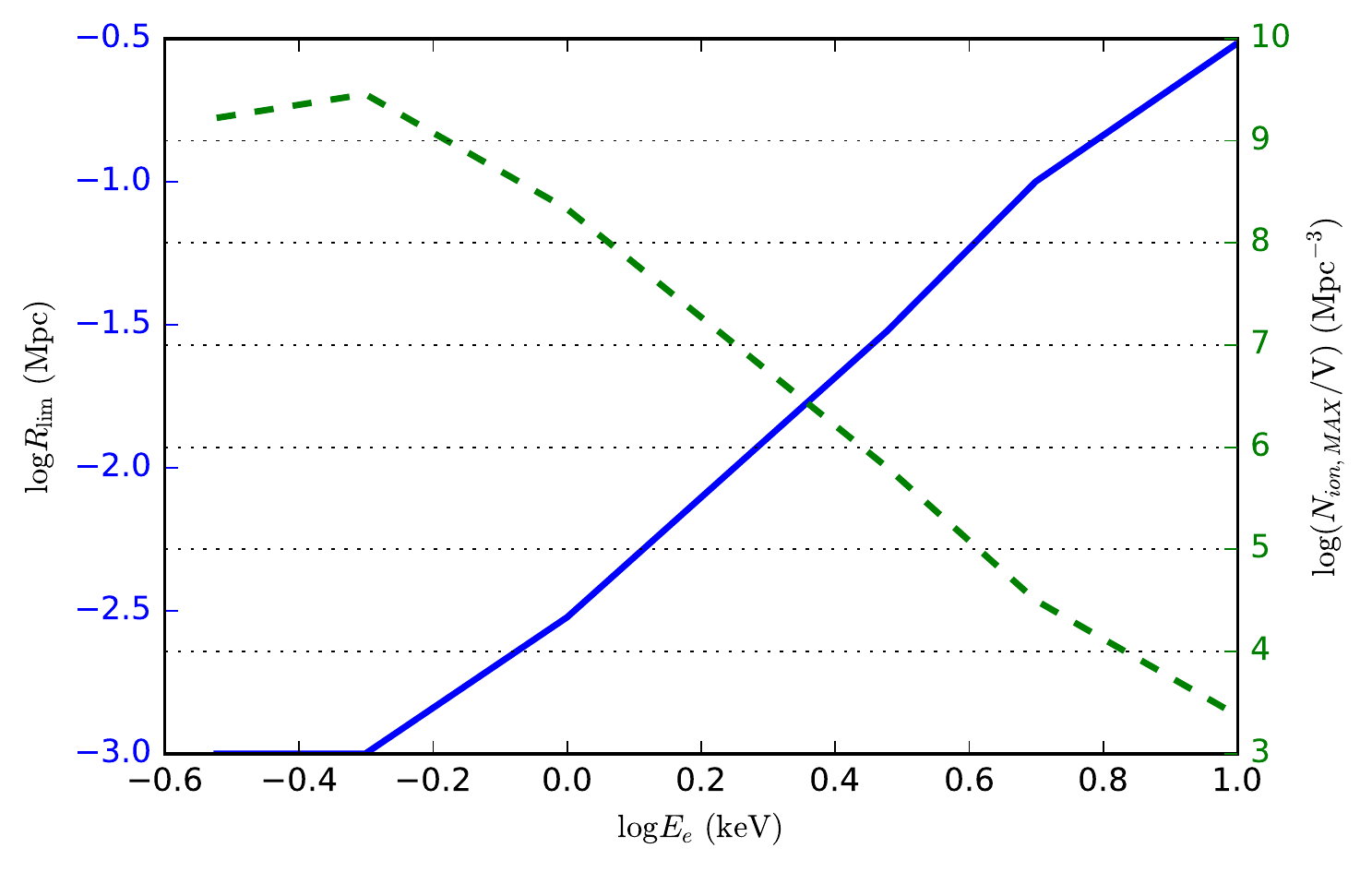}
\includegraphics[width=0.93\columnwidth]{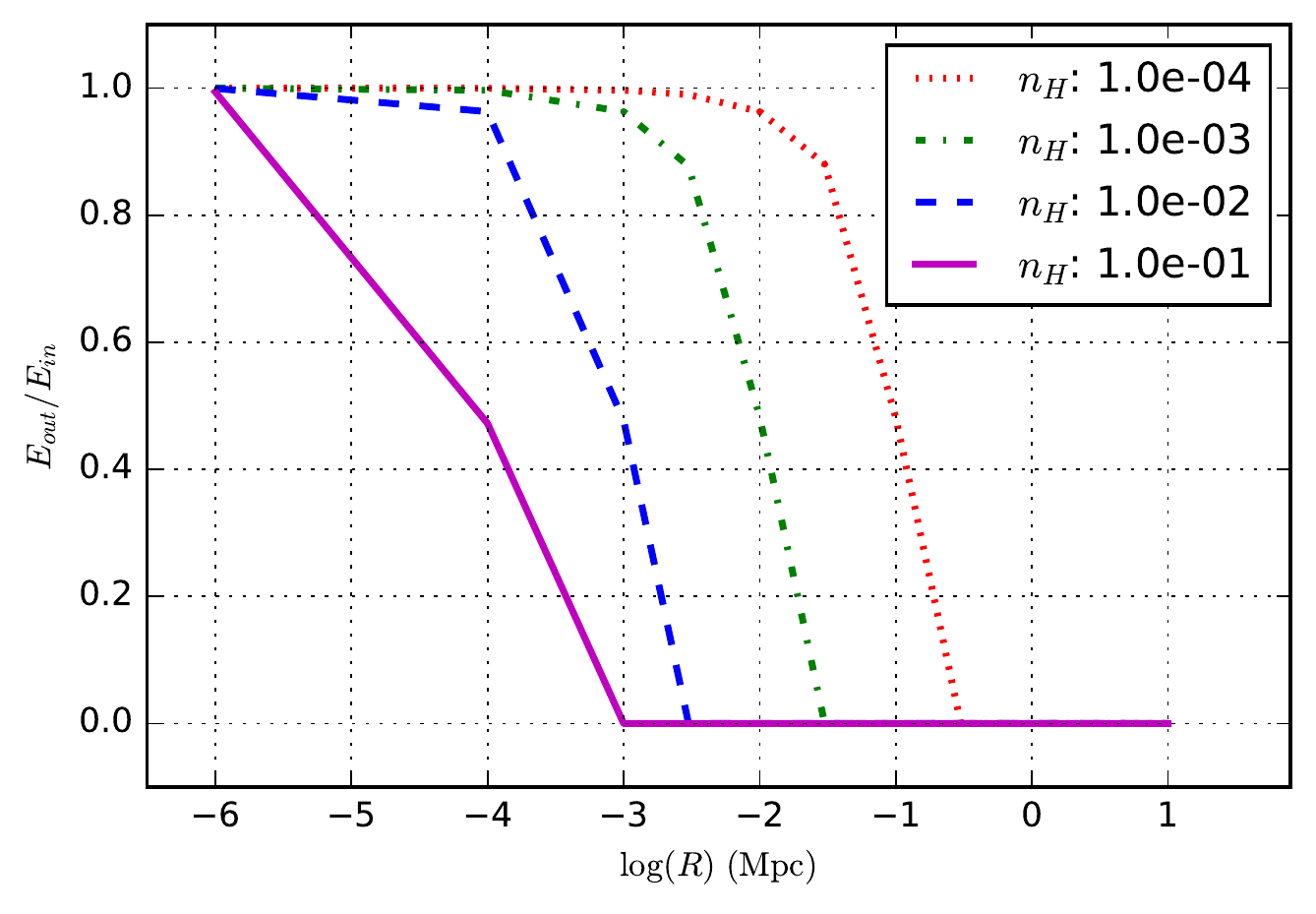}
\caption{{\em Upper panel:} Fraction $E_{\rm out}/E_{\rm in}$ of the energy injected
by the source, carried away by electrons as a function of the distance
$R$ to the source, for different primary electron energies. The medium is
homogeneous with a density typical of the IGM at redshift 10, and a fixed ionisation fraction ($f_{\rm ion} = 0.01$).
The radius $R_{\rm lim}$ at which no electron escapes delimits the sphere
accesible for ionisation and heating at each primary energy. 
{\em Central panel:} $R_{\rm lim}$ (solid blue line) and mean ionisation density per primary electron within $R_{\rm lim}$ (dashed green line), as a function of the kinetic energy of the primary electron. At fixed density and ionisation fraction, $R_{\rm lim}$ increases with primary energy but ionisations are smeared out in a larger volume, decreasing the mean ionisation density.
{\em Lower panel:} Same plot as the upper panel for different medium densities at a fixed primary energy of $10\,{\rm keV}$, showing the decrease of $R_{\rm lim}$ with increasing density. The hydrogen densities in the legend should be multiplied by the factor $(1+z)^3 \ \rm{cm}^{-3}$ with $z=10$.}
\label{fig:jet-IGMdensity:EoutvsR}
\end{figure}

In Fig.~\ref{fig:jet-IGMdensity:EoutvsR} (upper panel) we show the fraction of the energy injected by the source, carried away by the electrons that go across the surface of the sphere, for different energies of the primary electrons and at a fixed medium density and ionisation fraction. As it is expected, this fraction decreases with $R$ because energy is deposited in the medium as ionisation, heat and excitations (recombination is almost negligible in all cases). For each primary energy, there exists a maximum radius $R_{\rm lim}$ at which no electron escapes, because all the energy has already been deposited within it. This radius delimits the sphere accessible by electrons at a given energy, and can thus be called the ionisation (and heating) range. It increases with energy due to the increase of the mean free path, determined by the decrease of the cross sections of the relevant processes (Fig.~\ref{fig:jet-IGMdensity:EoutvsR}, central panel). For the same reason, $R_{\rm lim}$ increases with decreasing density, a behaviour clearly seen in the lower panel of the same figure. On the other hand $R_{\rm lim}$ does not vary significantly with $f_{\rm ion}$. Even though 10keV is the upper limit of this set of simulations, the aforementioned behaviour could be extrapolated to higher energies. As a consequence of their longer mean free path, high-energy electrons transport the energy farther away from the source than low-energy electrons, increasing $R_{\rm lim}$. Fig.~\ref{fig:jet-IGMdensity:EoutvsR} shows clearly that energy deposition is non-local. Once a low-energy electron is created, the ionisation and heat it produces is spread over kpc to Mpc.

These results have an important consequence for reionisation: at densities of
$0.1\,{\rm cm}^{-3}$ and above, typical of the ISM of a galaxy, even
$10\,{\rm keV}$ electrons travel $\lesssim 1\,{\rm kpc}$, which is of the order of magnitude of the size of a typical galaxy at $z \sim 10$ \citep{Bromm09}. In other words, electrons that have the largest probability of ionising, cool within the galaxy and cannot escape from it. Therefore, {\em to contribute to reionisation, low-energy electrons must be produced locally in the IGM, either by cooling of high-energy electrons or by other particle processes}. This highlights the need for {\em high-energy} CR sources such as MQs.

\begin{figure}
\includegraphics[width=\columnwidth]{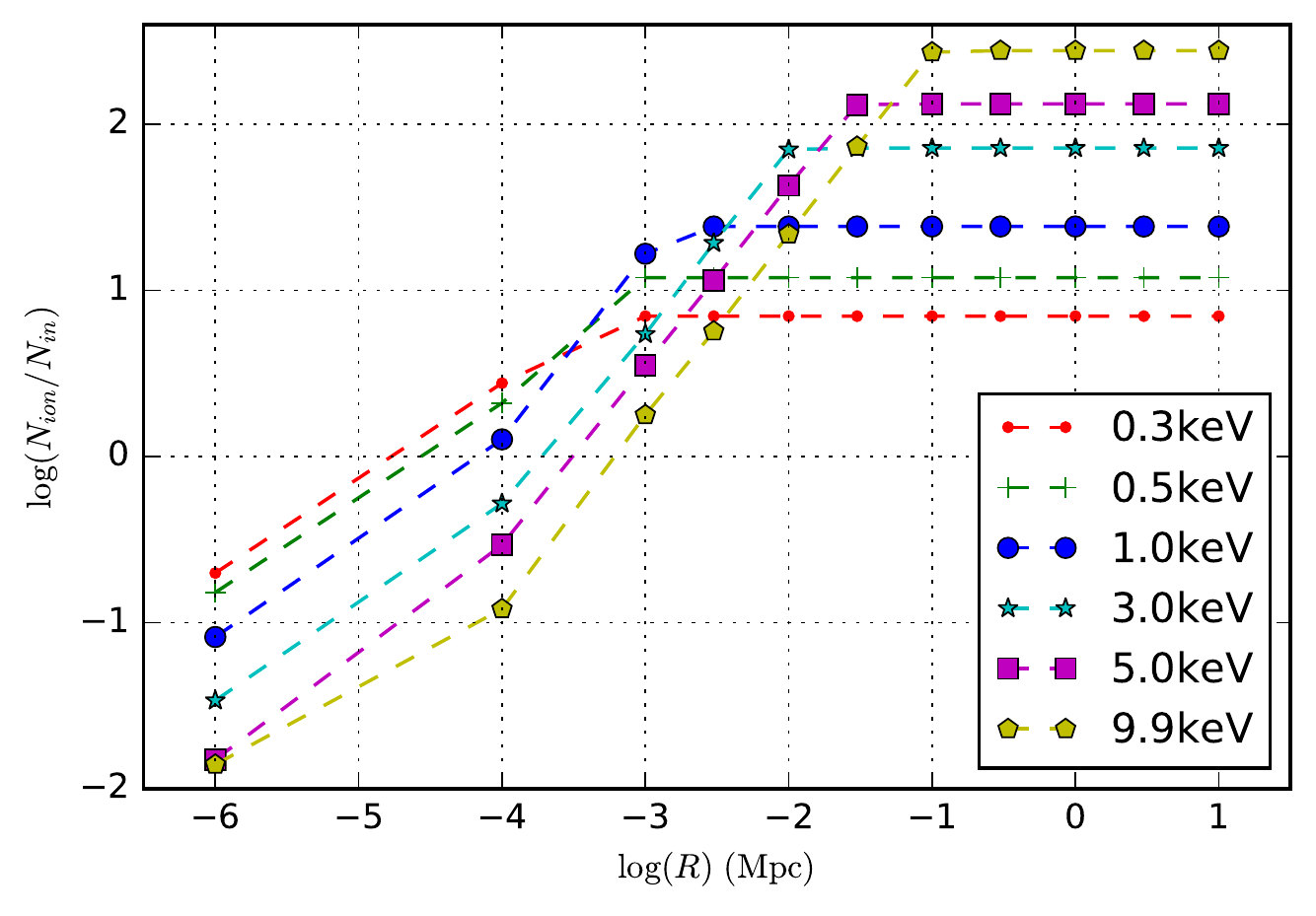}
\includegraphics[width=\columnwidth]{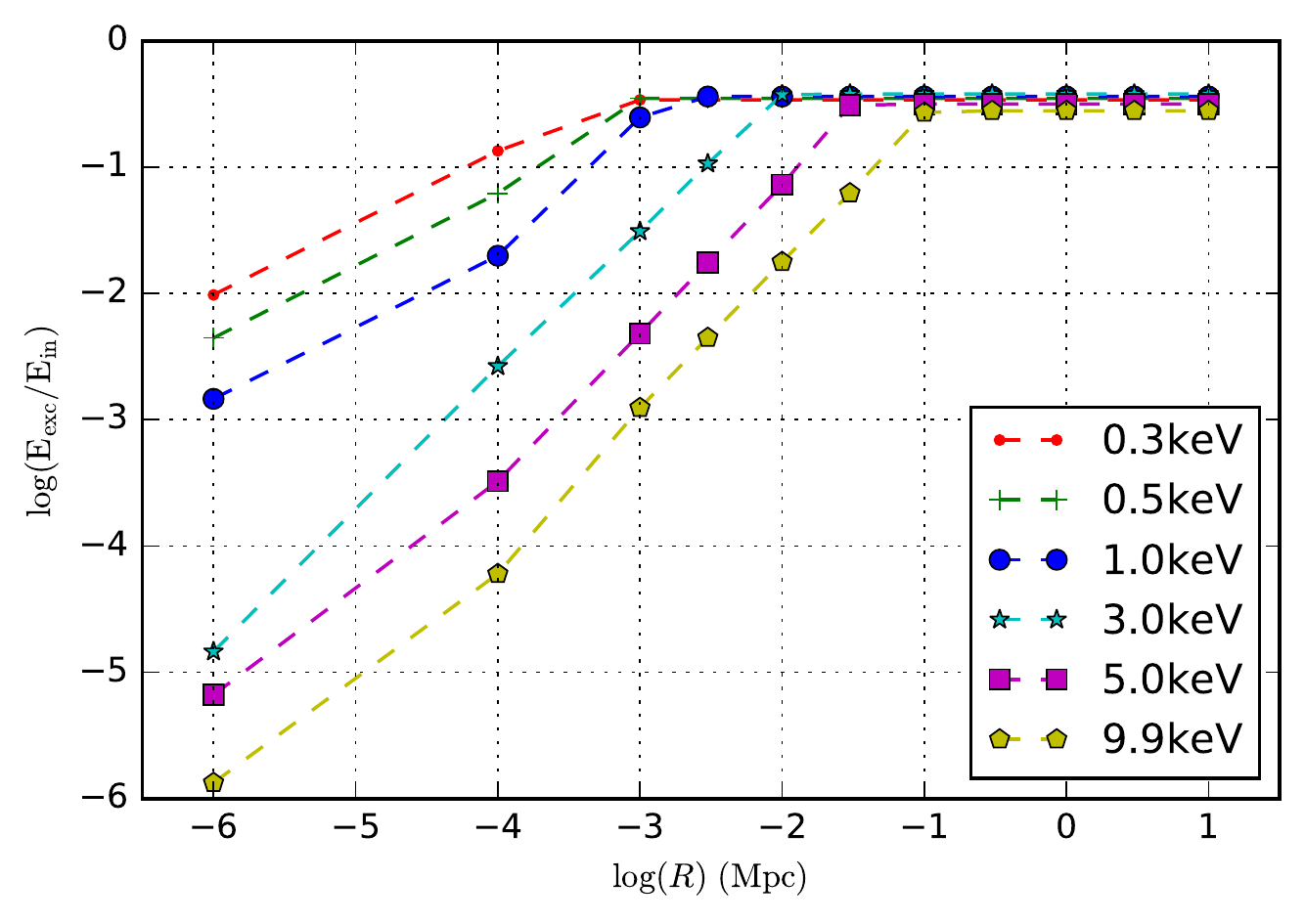}
\includegraphics[width=\columnwidth]{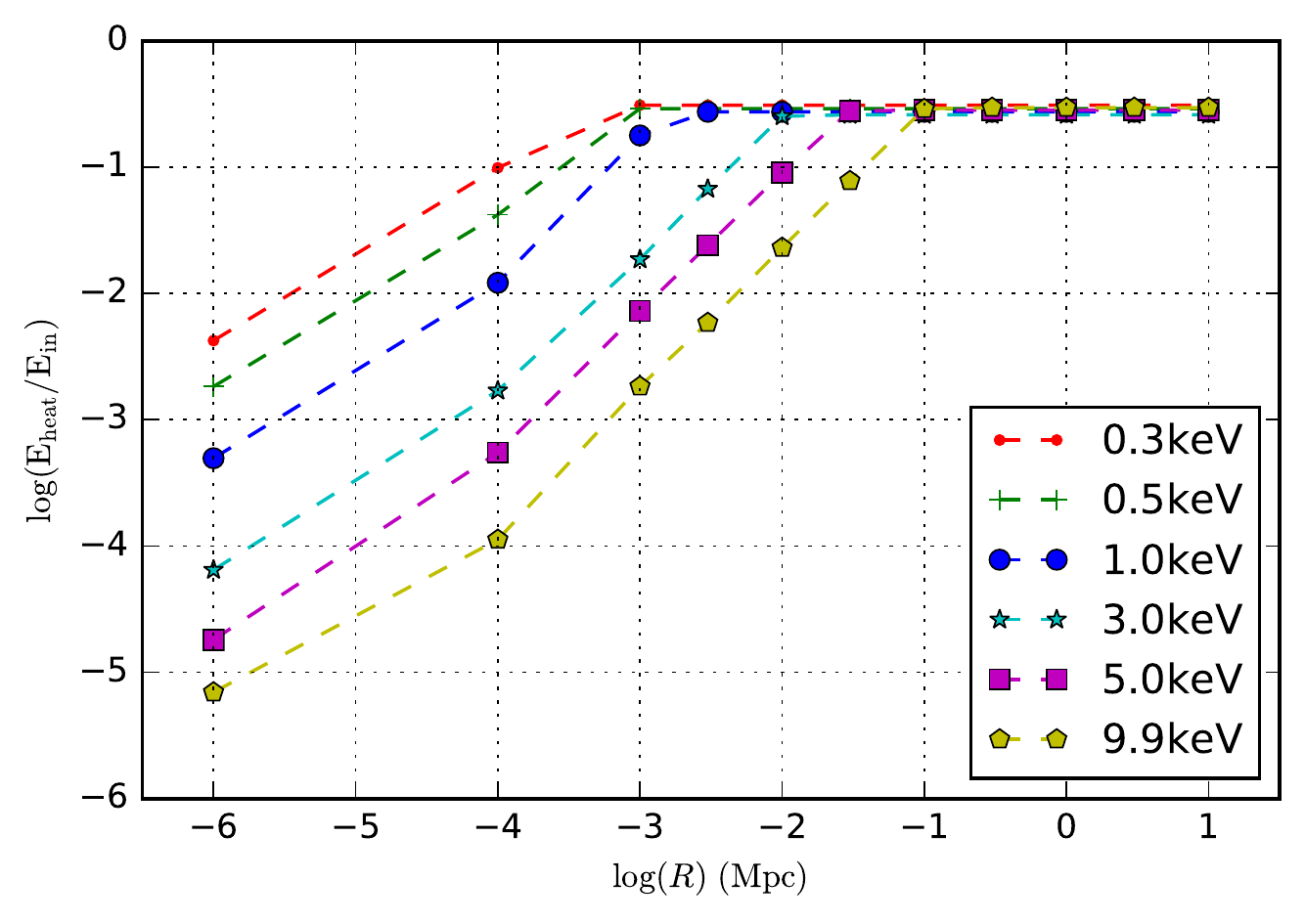}
\caption{Ionisation count per primary electron ($N_{\rm ion}/N_{\rm in}$, upper panel), excitation energy loss per unit of injected energy ($E_{\rm exc}/E_{\rm in}$, central panel), and heat produced per unit of injected energy ($E_{\rm heat}/E_{\rm in}$, lower panel), as a function of the distance $R$ from the source, for different kinetic energies $E_{\rm k}$ of the primary electrons. All curves grow with $R$ less steep than the volume, indicating that most of the energy deposition takes place close to the source. At a fixed radius ($R<R_{\rm{lim}}$), the three quantities decrease with energy, following the behaviour of the cross section. The total ionisation count increases with the electron energy.}
\label{fig:jet-IGMdensity:niovsR}
\end{figure}

Fig.~\ref{fig:jet-IGMdensity:niovsR} shows the simulated ionisation count ($N_{\rm ion}$, upper panel) per primary electron, the energy lost by excitation ($E_{\rm exc}$, central panel), and the heat ($E_{\rm heat}$, lower panel) produced within a sphere of radius $R$, the last two per unit of injected energy. All these quantities grow with radius until they reach a constant value at $R_{\rm lim}$, as expected. Their growth is, however, less steep than that of the enclosed volume, indicating that the largest effect on the medium is produced close to the source. 
For a fixed radius $R < R_{\rm lim}$, the amount of ionisations is larger for the less energetic electrons, following the behaviour of the cross section. The total number of ionisations within $R_{\rm lim}$ shows the opposite behaviour, increasing with electron energy because of the larger energy available for ionisation. However, the larger volume involved implies that the mean ionisation density produced within $R_{\rm lim}$ decreases with energy (Fig.~\ref{fig:jet-IGMdensity:EoutvsR}, central panel). A similar behaviour is obtained for the fraction of the energy transformed into heat and excitations. 

\begin{figure}
	\includegraphics[width=\columnwidth]{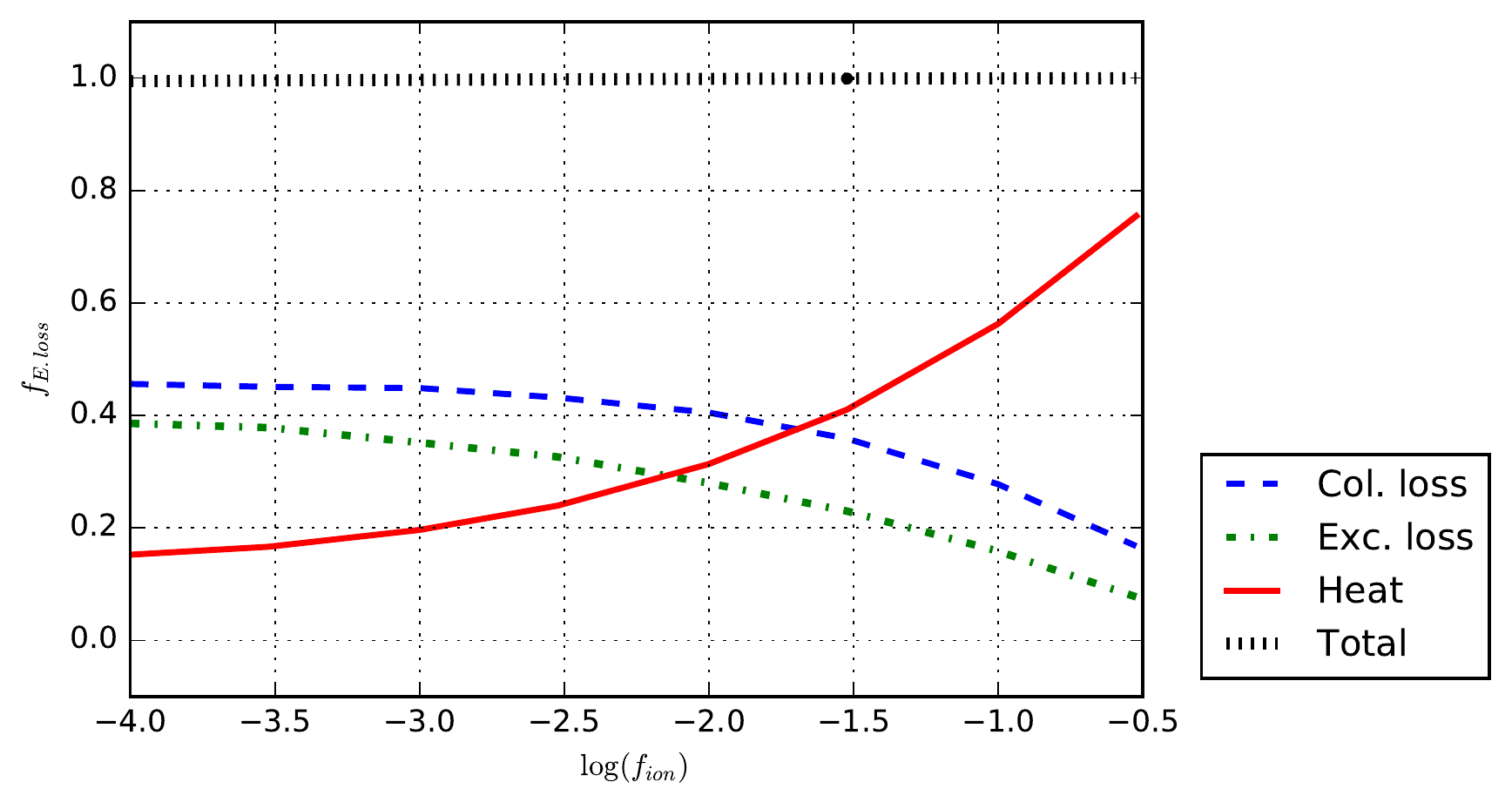}
    \caption{Fraction of the energy deposited in ionisations (dashed blue line), excitations (dashed-dotted green line) and heat (solid red line) for a $10\,{\rm keV}$ primary electron and a typical IGM density at $z = 10$. The electrons have been followed until they reached a distance of $10\, {\rm Mpc}$ from the source; hence, the escape fraction is null.}
\label{fig:jet-IGMdensity:Elossvsfion}
\end{figure}

For distances large enough that the electrons have ran out of energy, the fraction of the injected energy lost to the different processes depends mainly on the ionisation fraction of the medium, as we show in Fig.~\ref{fig:jet-IGMdensity:Elossvsfion}. This behaviour is known \citep[e.g.,][]{shull1985,valdesferrara2008,furlanettostoever2010,valdes2010}, and depends on the availability of free thermal electrons to drain energy from the primary electrons, which is converted to heat. Recombinations are found to be negligible. 

To test the performance of our {\em JET} code, its results have been compared with the work of \citet{valdesferrara2008}. {\em JET} reproduces fairly well the trends of the excitation, collisional ionisation, and heat as a function of the ionisation fraction of the medium, when the volume is large enough to comprise the whole energy deposition. The differences between our estimations and those of \citet{valdesferrara2008} for a 3 keV primary are typically around 3-5\%, reaching 10\% in the worst case. These small disagreements can be tracked to the differences in the approximations used for the cross sections of the basic physical processes. \citet{valdes2010} and \citet{furlanettostoever2010} have shown that a more accurate treatment of the basic physical processes results in some deviations from the results obtained by \citet{valdesferrara2008}. These are important at ionisation fractions near 1, a range that has not been included in our computations, which reach only $f_{\rm ion} = 0.1$. At moderate ionisation fractions the agreement is fair (differences up to $\sim 20\%$ at $f_{\rm ion} \sim 0.01$). This level of agreement is enough for our purposes, partly because the energies at which discrepancies arise contribute only marginally to the overall ionisation and heating rates, and partly because our main goal is to assess the importance of the contribution of MQs to IGM reionisation and heating. The latter can be determined in order of magnitude taking into account only the most relevant atomic processes \citep[HI ionisation and $e^-e^-$ scattering, e.g.,][]{furlanettostoever2010}, leaving a more accurate and detailed description for a future work in case our results show that the contribution is large enough to deserve further exploration.

The results from the present section imply that {\em an efficient mechanism for IGM reionisation driven by electrons, requires the existence of two components: some carrier that transports the energy outside the galaxies deep into the IGM; and some mechanism (cooling or reactions) that produces locally low-energy, high-ionising-power electrons from these carriers}. In our scenario, the carriers are high-energy electrons from MQ jets, and the cooling mechanism is provided by the interactions of the carriers with the CMB and the IGM itself.

\section{Escape of electrons from galaxies}
\label{sect:galaxy}

To explore the reionisation mechanism proposed above for MQs, we must set the appropriate scenario in the simulations. To simplify the problem, we model a galaxy at $z = 10$ as a homogeneous sphere of hydrogen. As low-energy electrons do not escape from the galaxy, and ionisation and excitations within the galaxy are of no interest for us, the detailed composition of the ISM is not relevant, as far as the total density is preserved. The pure-hydrogen sphere is used then to save computational time. The density of the ISM is set to a typical value of $1\,{\rm cm}^{-3}$. The ionisation fraction of the ISM ($f_{\rm ion}^{\rm ISM}$) and the radius of the galaxy ($R_{\rm g}$) are the free parameters of the model. The first one has been varied in the range $10^{-1}-10^{-4}$. Direct constraints on the galaxy size at $z = 10$ are not available, but the widely accepted hierarchical clustering scenario of structure formation and evolution \citep{White78,Fall80} indicates that galaxies should have been much smaller at these epochs than today. Indeed, it is assumed that primordial galaxies may look like dwarf star-forming galaxies at $z = 0$ \citep[e.g.,][]{Kunth00}. We adopt then three galaxy sizes for the simulation, $R_{\rm g} = 0.1, 0.5, 1\,{\rm kpc}$, which are of the order of magnitude of the sizes of present dwarf galaxies.

The model galaxy hosts MQ sources, located for simplicity at its center. The electron luminosity and spectrum of MQs is largely unknown. It depends on the jet composition (leptonic or hadronic), its launching mechanism, the acceleration and cooling processes within the jets, and in the termination shocks. All these problems are still unsolved, and different models have been proposed for each stage \citep[e.g.,][and references therein]{Romero14}. As pointed out by \citet{heinzsunyaev2002}, there may be a component of cold, low-energy electrons escaping from the bulk of the jet into the ISM. As MQ jets move at mildly relativistic Lorentz factors, these electrons should have kinetic energies of the order of MeVs or lower. On the other hand, synchrotron radio emission from jets, and $\gamma$-ray emission from MQs indicate the presence of ultrarelativistic electrons \citep{Soria2010} which would constitute a high-energy spectral component.

Given the lack of precise knowledge, we have chosen to model a broad simulated MQ spectrum. We adopt a low-energy cutoff of $10\,{\rm keV}$, because we have already shown that lower-energy electrons can not escape the galaxies (Sect.~\ref{subsect:sim-LE}). We set also a conservative high-energy cutoff of 1PeV to account for the possible existence of ultrarelativistic particles. In each simulation, we propagate with {\em UTOPIA} a set of 1000 electrons sampled from a uniform spectrum in $\log E_{\rm k}$. This allows us to have uniform statistics through the whole spectral range. Although real source (and particularly MQ) spectra are not flat in $\log E_{\rm k}$, the usefulness of our choice relies on the non-existence of interactions between the particles in the energy cascade (they only interact with photon and matter fields). In this case all the results of the simulations are linear on the primary spectrum, and the Monte Carlo samples obtained from our simulation can be appropriately weighted {\em a posteriori} to obtain the results for any source spectrum, as far as its cutoffs remain within the simulated energy range. For the same reason, all results are linear in the total kinetic luminosity of electrons, therefore we adopt a fiducial value $L_{\rm k} = 10^{40}\,{\rm erg\,s}^{-1}$. We stress that in our model the simulated luminosity and spectrum represent the integrated emission of the whole population of MQs in the galaxy, which could contain a luminous single source or a set of less-luminous ones. However, it is interesting to point out that the star formation rate of typical galaxies at $z=10$ is similar to that of local low-metallicity dwarfs \citep{madaufragos2017}, which contain at most a few XRBs \citep[][and references therein]{douna2015}. In any case, the linearity of the results on $L_{\rm k}$ ensures that they can be conveniently scaled to any total luminosity value. 

\begin{figure}
\includegraphics[width=\columnwidth]{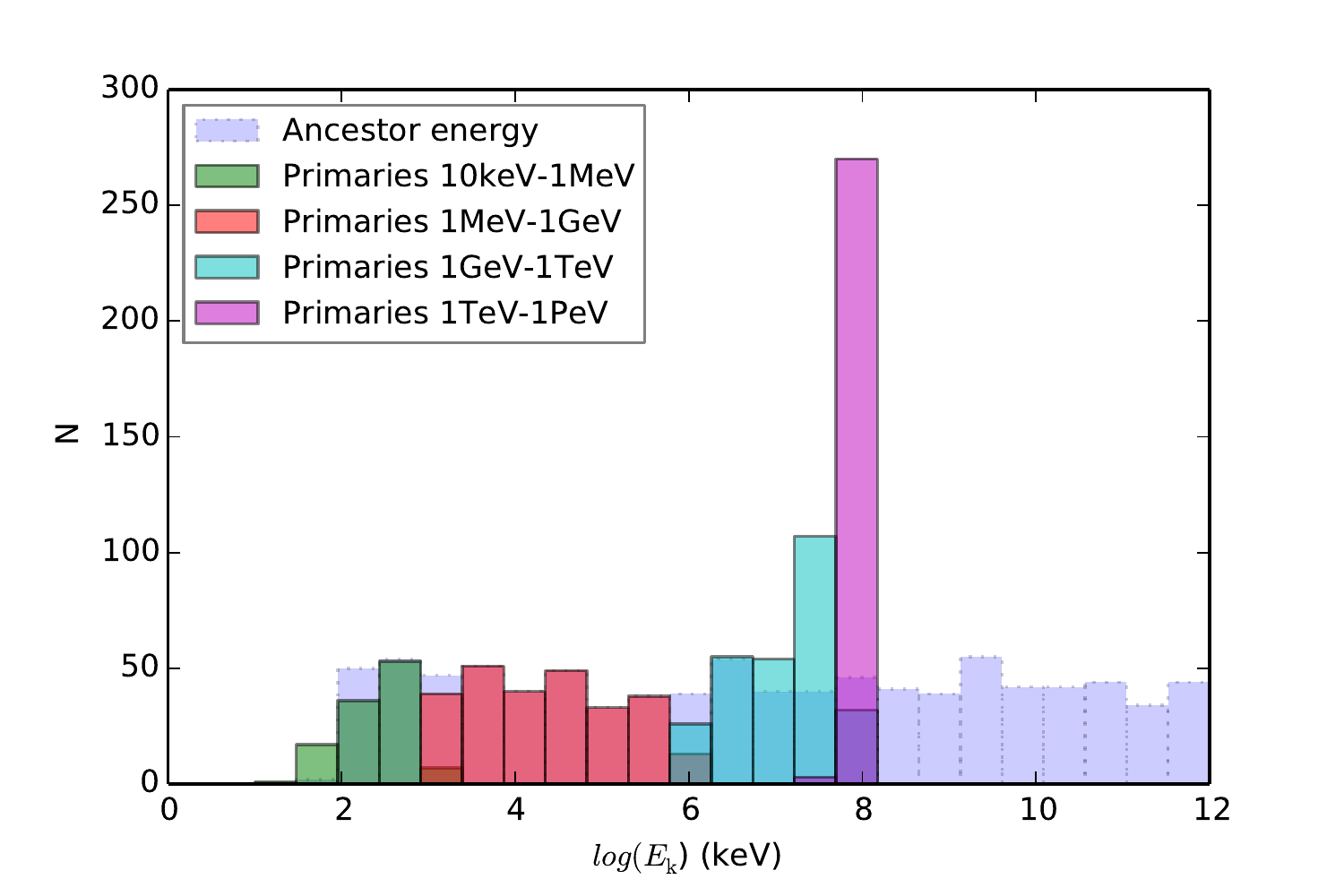}
\caption{Kinetic energy spectrum of the electrons that escape the galaxy, originally injected with kinetic energies in the $10\,{\rm keV}-1\,{\rm PeV}$ range. The different colors of the main histogram show the range of kinetic energies of the primaries. The light-purple (shadowed) histogram shows the number of outcoming particles as a function of the energy of the primary electron that gave birth to them (ancestor), for comparison. The histogram corresponds to 1000 injected electrons and a galaxy with $R_{\rm g} = 1\,{\rm kpc}$ and $f_{\rm ion} = 0.01$.}
\label{fig:UTOPIA-gxy:histsalidaconancestros}
\end{figure}

In order to characterise the energy loss inside our model galaxy, we study the spectrum of the electrons escaping from it (Fig.~\ref{fig:UTOPIA-gxy:histsalidaconancestros}). The different colors in this figure correspond to the range of energy of the primary electron that gives birth to the escaping electrons. In the low energy limit, we see that just a few electrons that were originally emitted with kinetic energies in the $30-100\,{\rm keV}$ range come out of the galaxy. This is due to the fact that in this energy range ionisations and heat are still effective, therefore most of the energy is deposited inside the galaxy. The amount of electrons that escape the galaxy with $10\,{\rm keV}-1\,{\rm MeV}$ increases as the energy grows, and they mostly correspond to primaries in the same range. In the MeV--GeV range, most electrons emitted lose just a small fraction ($\sim 1\%$) of their kinetic energies while escaping the galaxy. This is because the main cooling mechanism is IC proceeding in the Thomson limit (other processes are negligible). The contribution of IC cooling channel grows from $\sim 1\%$ of the energy loss at $1\,{\rm GeV}$ to $\sim 95\%$ at $1\,{\rm TeV}$, as the escaping electron energy falls from $\sim 98\%$ to $\sim 5\%$. All the electrons that were originally emitted with energies in the $1\,{\rm TeV} - 1\, {\rm PeV}$ range, come out of the galaxy with energies $\lesssim 1\,{\rm TeV}$. This is due to the fact that at kinetic energies of the electron around $20-30\,{\rm TeV}$, IC against the CMB enters the Klein Nishina regime, which implies that in each collision they transfer a large amount of their energies to CMB photons. As a by-product, this cooling generates a large population of high-energy $\sim {\rm GeV}$ photons. These results do not depend on the ionisation fraction of the ISM.

From the presented toy model, it is clear that a galaxy that contains sources of relativistic electrons turns out to be also a source of high- and low-energy electrons, that are injected into the IGM. The characteristic spectrum of a source accelerating electrons, such as a MQ, is generally described by a power law $N(E) \propto E^{-\alpha}$, where $E$ is the total electron energy. This is of course different from a uniform spectrum in $\log E_{\rm k}$, which has been previously used for the sake of simplicity. The power law results from the Fermi acceleration processes, and typically for MQs can be modelled by an index $\alpha = 2.5$. As stated above, the previous results could be weighted according to the real energy distribution of the source. In any case, power law spectra generally involve a higher proportion of low energy electrons, which have been shown to be better ionising agents than their high energy counterparts. 

\begin{figure}
\includegraphics[width=\columnwidth]{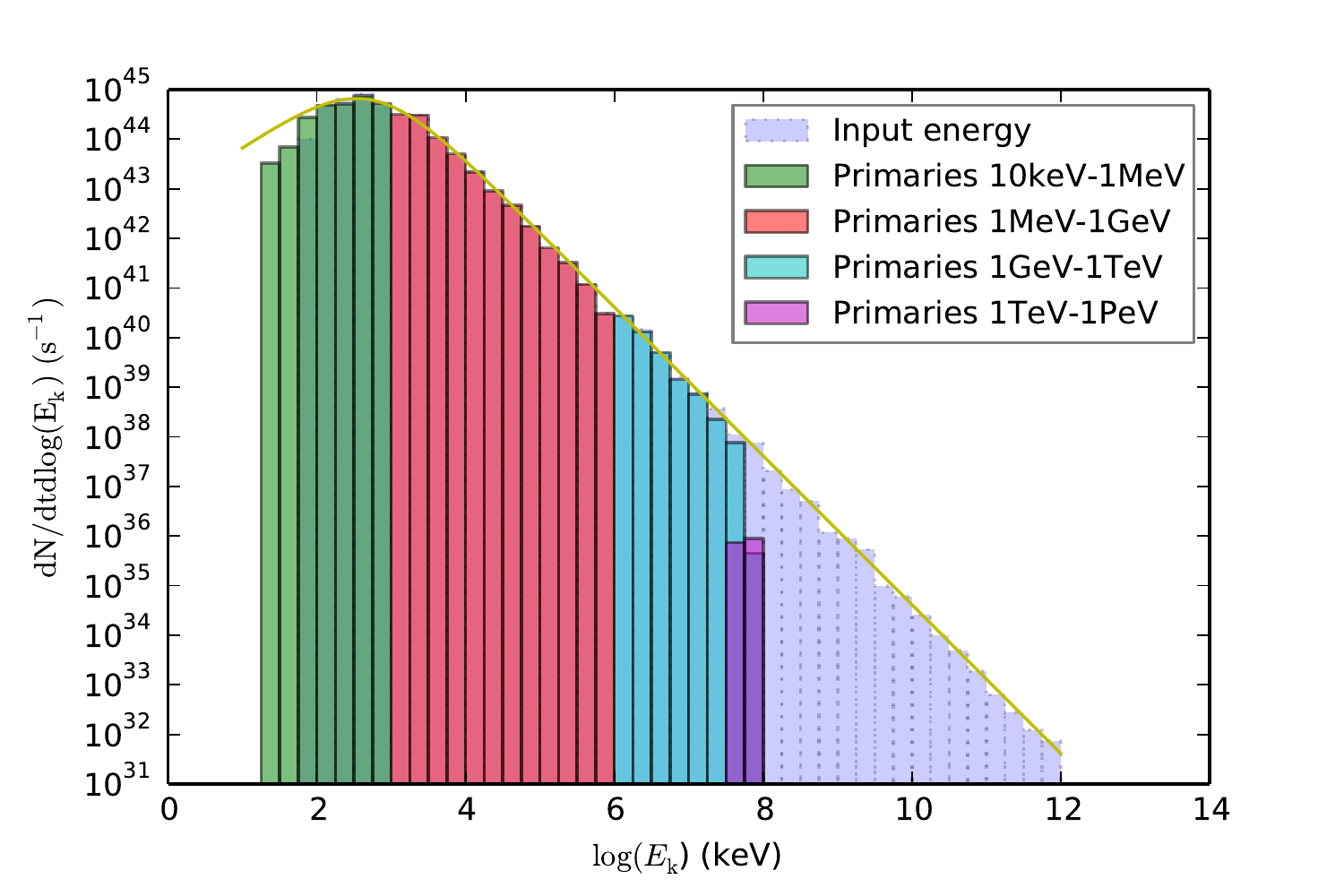}
\caption{Spectral distribution of electrons escaping the galaxy, injected by a MQ source with a total luminosity of $10^{40}\, {\rm erg\,s}^{-1}$ in a galaxy with $R_{\rm g}=1 \ \rm{kpc}$ and $f_{\rm ion}=0.01$. Colours represent the kinetic energies of the primary electrons ejected by the MQ. The solid line is the original spectrum of the source. The light-purple (shadowed) histogram is the same as in Fig. \ref{fig:UTOPIA-gxy:histsalidaconancestros}. Ionisation and elastic scattering cool electrons in the low-energy end of the spectrum, whereas IC is responsible for depleting the high-energy end.}
\label{fig:UTOPIA-gxy:histsalidaconvpowlaw}
\end{figure}

To obtain an example of a typical energy distribution of escaping electrons for a population of MQ sources, we have weighted the simulated spectrum using for the primary spectrum, a power law between $E = 521\,{\rm keV}$ ($E_{\rm k} = 10\,{\rm keV}$) and $1\,{\rm PeV}$, with an index of $2.5$. A total kinetic luminosity of $10^{40} {\rm erg\,s}^{-1}$ was adopted for the MQ source. In Fig.~\ref{fig:UTOPIA-gxy:histsalidaconvpowlaw}, we show the spectrum of the electrons that escape the galaxy in this case.

In line with the previously described uniform input spectrum, in the low-energy limit many electrons do not escape the galaxy, due to the high ionisation cross section and the considerable amount of energy deposited as heat in the medium. In the $1\,{\rm MeV}-0.1\,{\rm TeV}$ range, the spectrum of the outcoming electrons corresponds to a power law, that matches the input electron spectrum. For kinetic energies higher than a few dozens of TeVs, electrons lose their energy by means of IC cooling in the Klein-Nishina regime, giving birth to a large amount of photons and an increased population of $\lesssim 1\, {\rm TeV}$ electrons. However, this excess is not perceptible in the weighted spectrum due to the steep power-law spectrum of the source. As a consequence of the efficiency of this mechanism, a lack of electrons with energies in the $1\,{\rm TeV}-1\,{\rm PeV}$ range is seen in the outcoming spectrum.

\section{IGM ionisation and heating by microquasars}
\label{sect:igm}

In order to explore the effect on the IGM of the electrons escaping the galaxies, we include in our scenario a low-density medium surrounding the model galaxies. Its density is $n_{\rm IGM} = 2.4 \times 10^{-4}\,{\rm cm}^{-3}$ (the mean baryonic density of the Universe at $z = 10$), whereas its ionisation fraction $f_{\rm ion}^{\rm IGM}$ is a free parameter. We have set the ionisation fraction of the ISM arbitrarily to $0.01$, because we have shown in the previous section that this parameter does not change significantly the escaping electron spectrum. We assume that the source is a single MQ (or a population of them) with a power law primary spectrum and a total kinetic luminosity of $10^{40}\, {\rm erg\,s}^{-1}$, as in the previous section. We recall that all the resulting spectra are linear in this parameter. The spectral index has been varied within the typical range for MQs ($\alpha \in [2,3]$), and the low- and high-energy cutoffs have been varied freely within the whole energy range ($10\,{\rm keV}-1\,{\rm PeV}$). We compare all our results to our fiducial model defined by $\alpha = 2.5$ and energy cutoffs equal to the corresponding limits of the simulated source spectrum.

We have followed the propagation of the electrons with {\em UTOPIA}, and have studied the spatial variation of their energy deposition from the outer radius of our model galaxy, until they reach a distance of $1\,{\rm Mpc}$ from its center. This value has been adopted because it is presumed to be the typical galaxy separation at $z \sim 10$ \citep[see][]{madaufragos2017}. As {\em UTOPIA} follows electrons only down to $E_{\rm k} = 3\,{\rm keV}$, we have complemented our analysis using the results of {\em JET} (Fig.~\ref{fig:jet-IGMdensity:niovsR}) to add, for each particle cooled down below this limit, the number of ionisations and the heat deposited until it thermalises.

\subsection{Ionisation}

\begin{figure}
\includegraphics[width=\columnwidth]{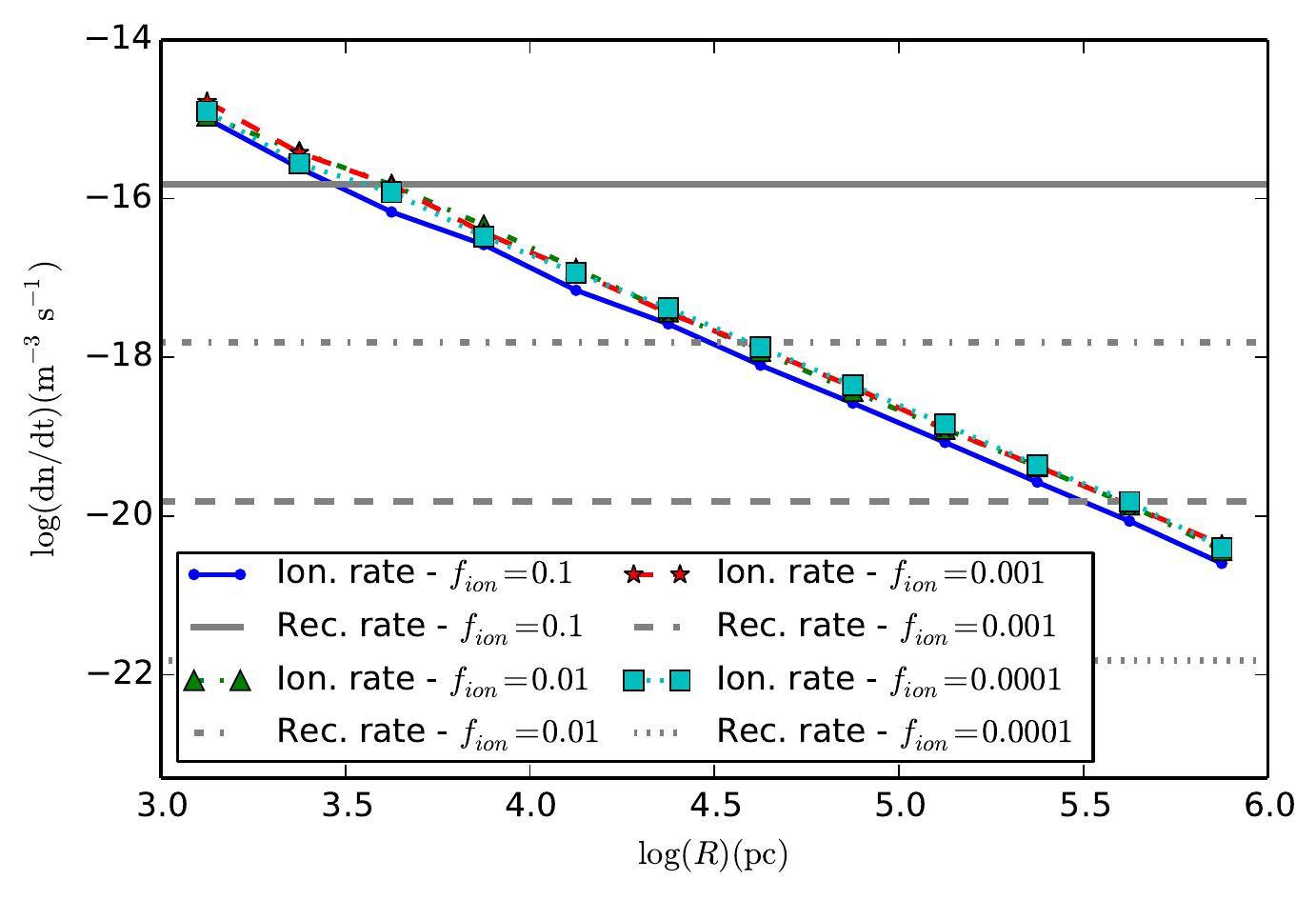}
\caption{Ionisation and recombination rate densities as a function of the distance from the centre of a galaxy at $z = 10$ ($R_{\rm g} = 1\,{\rm kpc}$), for different ionisation fractions of the IGM, and for a MQ source with a power-law spectrum of index $2.5$ an a kinetic luminosity $L_{\rm k} = 10^{40}\,{\rm erg\,s}^{-1}$. Ionisation rate densities are almost independent of the ionisation fracction of the IGM.}
\label{fig:UTOPIA-IGMion:nion-nrec_radius}
\end{figure}

Fig.~\ref{fig:UTOPIA-IGMion:nion-nrec_radius} shows the ionisation rate per unit volume obtained for our fiducial model, compared to the recombination rate (calculated from case B recombination coefficient $\alpha_B\approx 2.6 \times 10^{-13} \rm{cm}^3 \rm{s}^{-1}$), for different ionisation fractions of the IGM. As it is expected, the ionisation rate per unit volume decreases as the distance from the source increases. It shows also a very weak dependence on the ionisation fraction of the IGM. Given that the recombination rate is proportional to the square of this fraction, the ability of MQ electrons to ionise the IGM at a fixed distance from the galaxy centre decreases with increasing $f_{\rm ion}^{\rm IGM}$. An estimate of the maximum $f_{\rm ion}^{\rm IGM}$ that can be attained at a fixed distance from the source is roughly given by the intersection of the curves. According to our results, a source emitting electrons as our fiducial model would ionise the IGM up a to sizeable value $f_{\rm ion}^{\rm IGM} = 0.1$ to within some kiloparsecs of the galaxy.

It is important to point out that the recombination rate used in Fig. \ref{fig:UTOPIA-IGMion:nion-nrec_radius} corresponds to a temperature $T_{\rm{IGM}}\sim10^4 \rm{K}$, which assumes a pre-heated medium (``warm'' scenario). This assumption is consistent with lower-redshift observational constraints on the thermal history of the IGM \citep[for example][]{theuns2002,zaldarriaga2001} and with many reionisation scenarios that rely on a pre-heated intergalactic medium, for example, some schemes based on the effect of X ray emission \citep[for example][]{ricottiostriker2004,mirabel2011,tanakapernahaiman2012} . However, there are many uncertainties related to the nature and relative importance of the sources of the thermal input, and their evolution with redshift. A colder medium would imply a higher recombination rate, and thus a lower reionisation power of electrons. 

\begin{figure}
\includegraphics[width=\columnwidth]{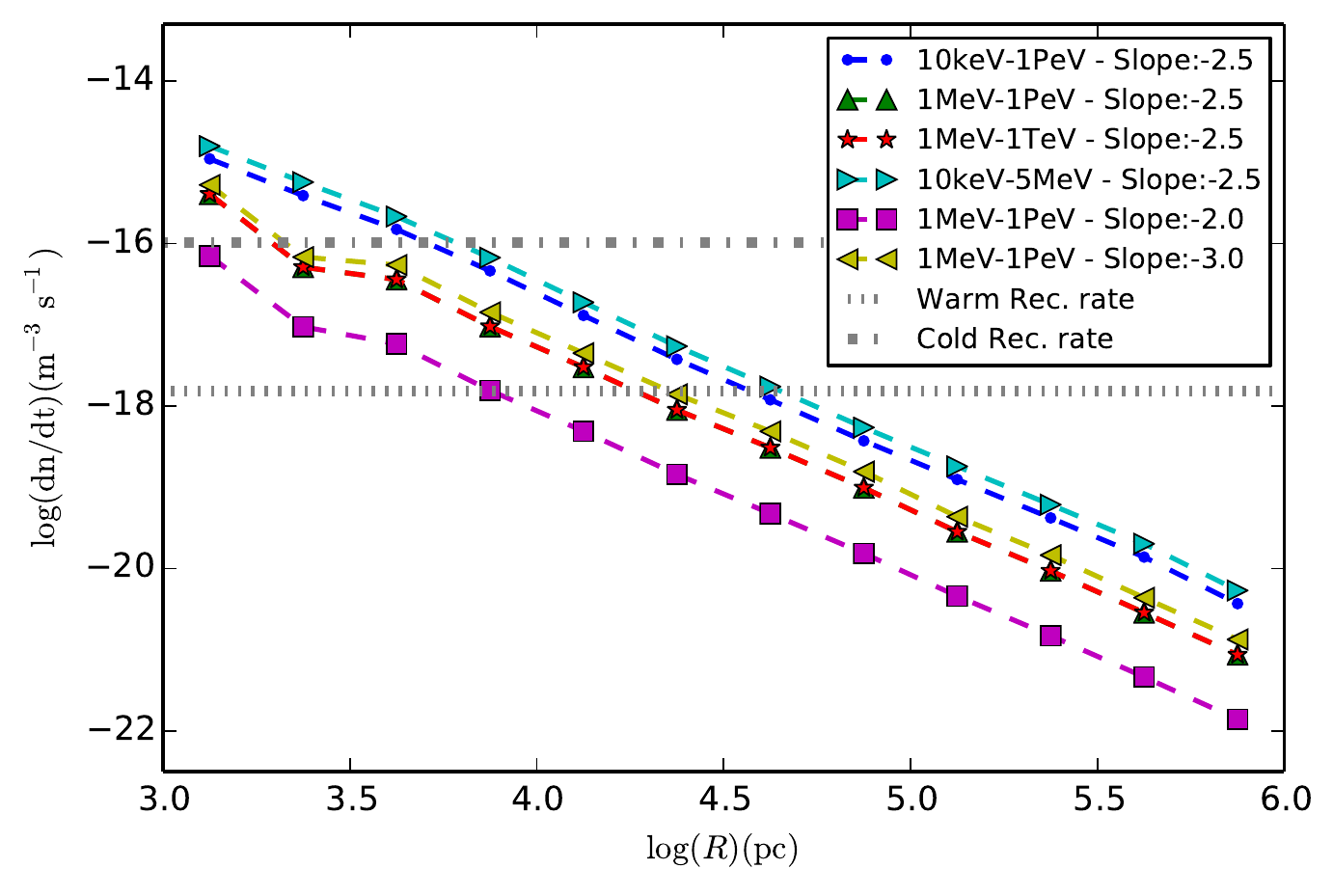}
\caption{Ionisation and recombination rate densities  as a function of the distance from the centre of a galaxy at $z = 10$ ($R_{\rm g} = 1\,{\rm kpc}$), for a fixed $f_{\rm ion} = 0.01$, and different electron spectra of the MQ source (lower and upper energy limits and slope of the power law) with a kinetic luminosity $L_{\rm k} = 10^{40}\,{\rm erg\,s}^{-1}$. Spectra displaying larger fractions of low-energy electrons produce larger ionisation rates. For comparison, the recombination rates were calculated for a warm recombination scenario, corresponding to a pre-heated medium with an IGM temperature $\sim 10^4$ K, and an extremely cold recombination scenario, corresponding to an IGM temperature equal to that of the CMB at $z=10$.}
\label{fig:UTOPIA-IGMion:nion-nrec_radius-parameters}
\end{figure}

Fig.~\ref{fig:UTOPIA-IGMion:nion-nrec_radius-parameters} shows the ionisation rates per unit volume for different primary electron spectra. It is clear from the observed trends that the ionisation rate does not depend on the high-energy cutoff of the spectrum. This is due to the fact that high-energy electrons contribute with a negligible fraction of the ionisations in the IGM. Contrary to this, the ionisation rate is sensitive to the low-energy cutoff because of the key role played by low-energy electrons. On the other hand, models with flatter spectra predict lower ionisation rates, which is consistent with a smaller ratio of low-to-high energy particles. 

For comparison, in Fig.~\ref{fig:UTOPIA-IGMion:nion-nrec_radius-parameters} an upper bound for the  recombination rate was calculated considering an extremely ``cold'' scenario, in which the temperature of the IGM was chosen to be equal to the CMB temperature at redshift $z=10$ ($T\sim 30\rm{K}$). The total recombination coefficient at this temperature was calculated from \citet{vernerferland1996}. This coefficient is around a hundred times higher than the case B recombination coefficient which causes an equal increase of the recombination rate. The ionizing power of the electrons at a fixed distance from the modelled source is at least the one defined by the difference between the ionisation and recombination rate in the ``cold'' model. 

\begin{figure}
\includegraphics[width=\columnwidth]{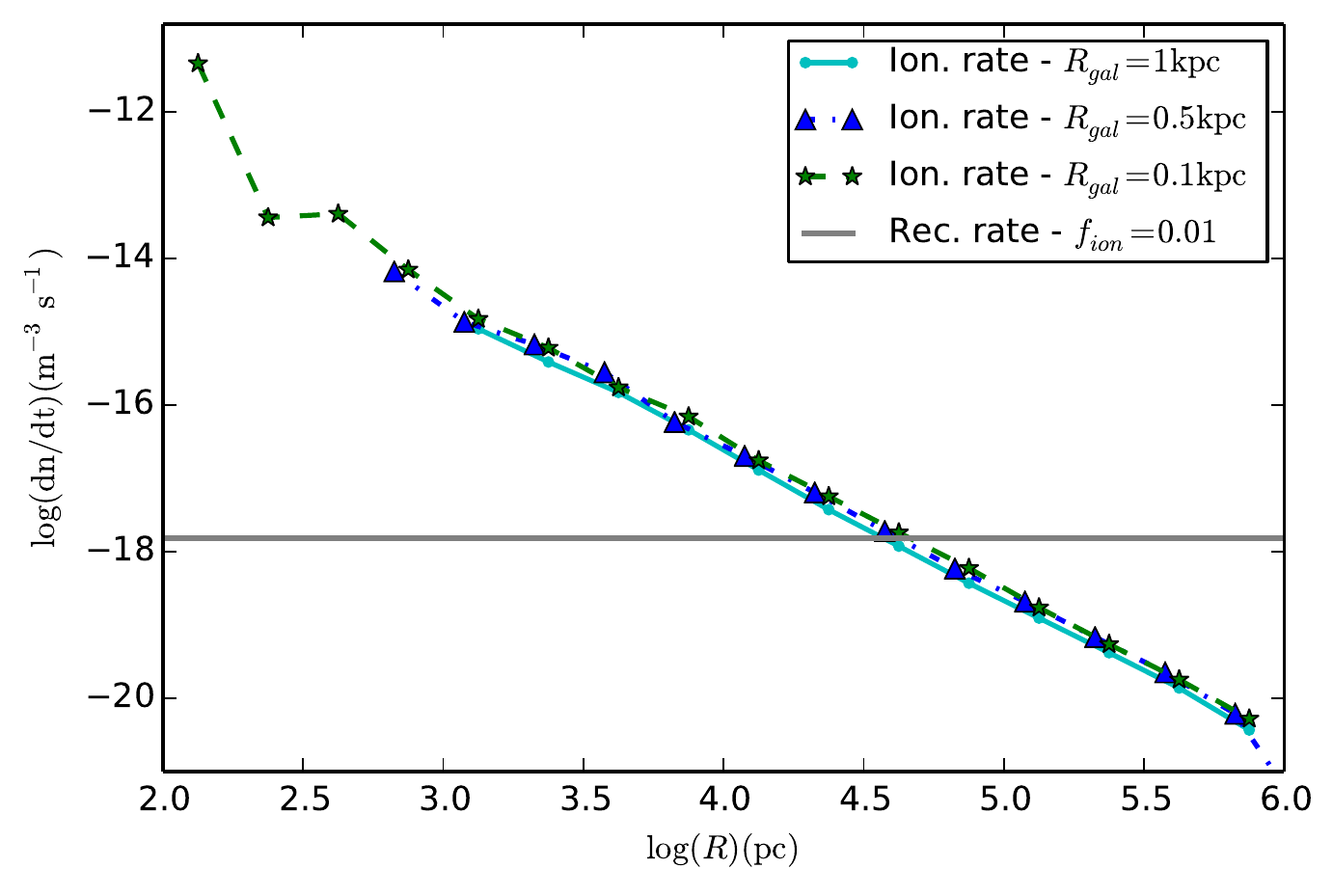}
\caption{Ionisation and recombination rate densities (warm scenario) as a function of the distance from the centre of a galaxy at $z = 10$, for different galaxy radii, a fixed $f_{\rm ion} = 0.01$, and our fiducial spectrum of a MQ source with a kinetic luminosity $L_{\rm k} = 10^{40}\,{\rm erg\,s}^{-1}$. Small galaxies are slightly better ionising sources, because more low-energy electrons can escape from them.}
\label{fig:UTOPIA-IGMion:nion-nrec_radiusgal}
\end{figure}

In Fig.~\ref{fig:UTOPIA-IGMion:nion-nrec_radiusgal} we present the ionisation rate per unit volume for different radii of the model galaxy. It increases slightly as the galactic radius is decreased. This is due to the fact that a larger number of low-energy electrons can escape from smaller galaxies. Once again, the ionisation rates decreases as a function of the distance from the source, therefore the highest ionisation rate is reached for the smallest galaxy.

\subsection{Heating}

\begin{figure}
\includegraphics[width=\columnwidth]{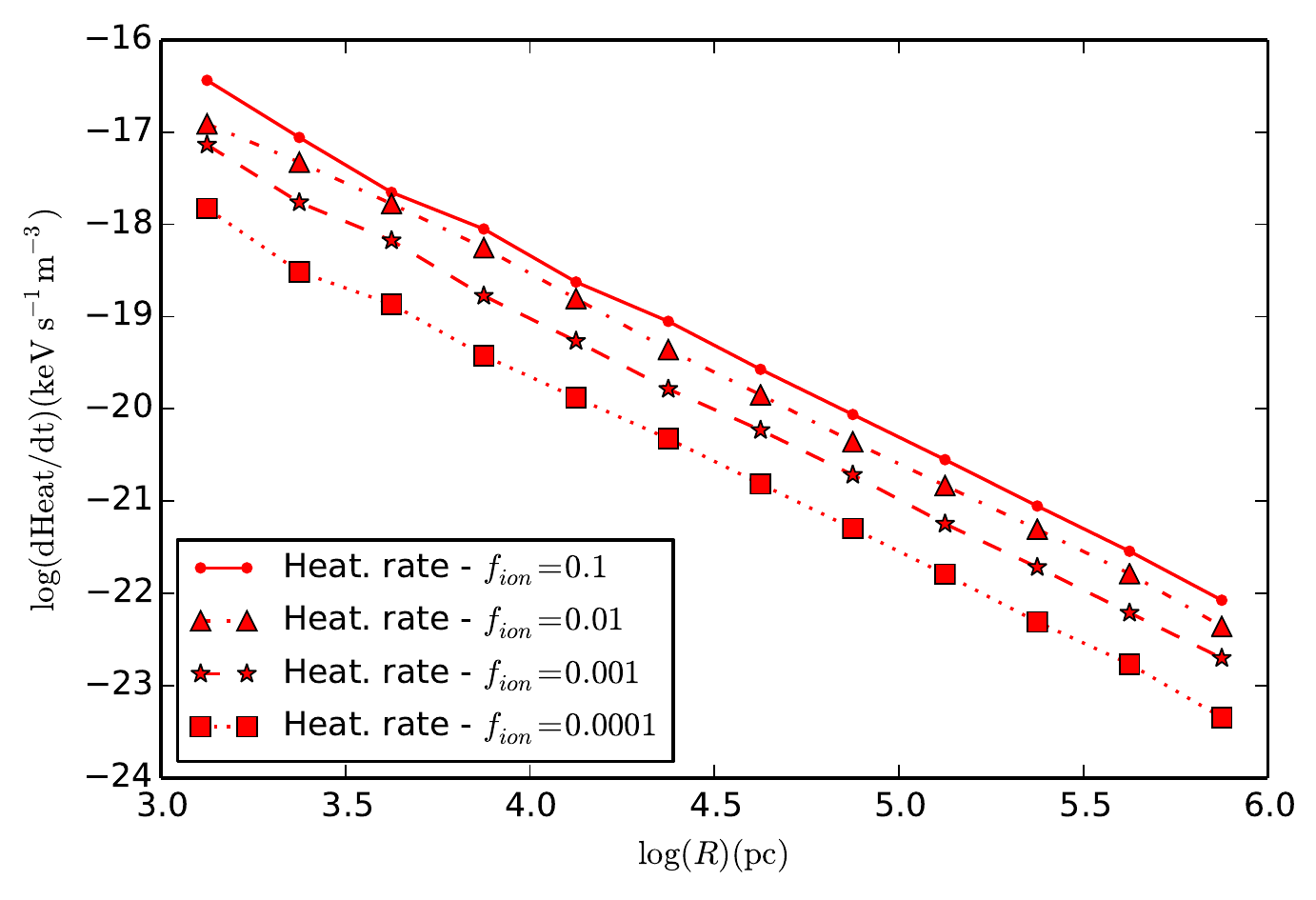}
\caption{Heating rate density as a function of the distance from the centre of a galaxy at $z = 10$ ($R_{\rm g} = 1\,{\rm kpc}$), for different ionisation fractions of the IGM, and for a MQ source with a power-law spectrum of index $2.5$ an a kinetic luminosity $L_{\rm k} = 10^{40}\,{\rm erg\,s}^{-1}$. Heating is strongly dependent on the ionisation fraction of the IGM, because it is mediated by elastic electron scattering.}
\label{fig:UTOPIA-IGMion:tempIGM_radius}
\end{figure}

In Fig.~\ref{fig:UTOPIA-IGMion:tempIGM_radius}, the heating rate density at $z = 10$ predicted by our models is shown as a function of the distance to the centre of the galaxy, for different ionisation fractions of the IGM. The behaviour is similar to that of the ionisation rate density, decreasing as the distance from the galaxy increases. In this case, however, the heating rate density of the IGM is strongly dependent on the ionisation fraction, increasing as the latter increases, because heat deposition is mediated by scattering with free electrons.
 
\begin{figure}
\includegraphics[width=\columnwidth]{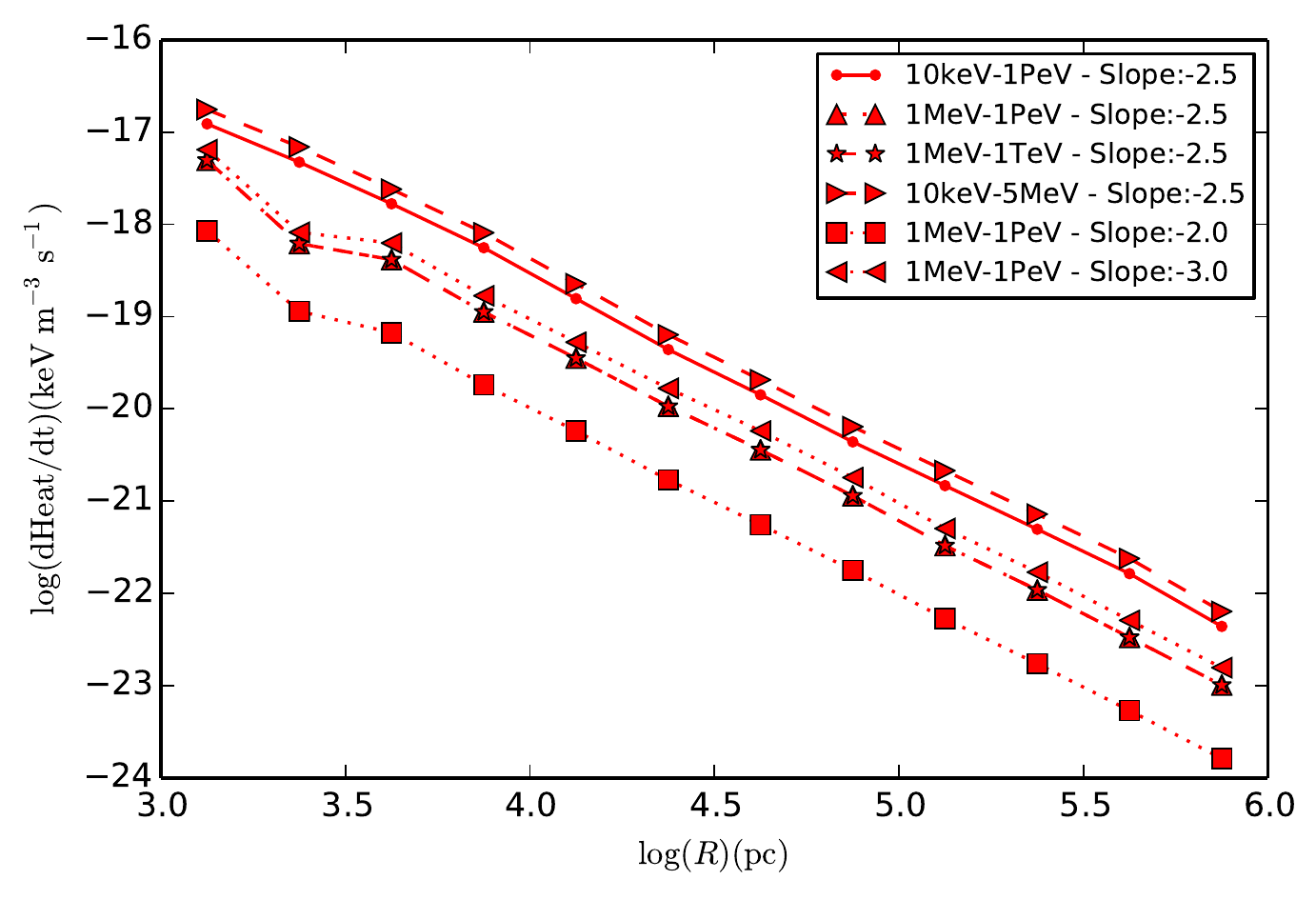}
\caption{Heating rate density as a function of the distance from the centre of a galaxy at $z = 10$ ($R_{\rm g} = 1\,{\rm kpc}$), for a fixed $f_{\rm ion} = 0.01$, and different electron spectra of the MQ source (lower and upper energy limits and slope of the power law) with a kinetic luminosity $L_{\rm k} = 10^{40}\,{\rm erg\,s}^{-1}$. Spectra displaying larger fractions of low-energy electrons produce larger heating rates.}
\label{fig:UTOPIA-IGMion:heat_radius-parameters}
\end{figure}

Regarding the parameters of the MQ electron spectrum, the heating rate density behaves in identical way as the ionisation rate density (Fig.~\ref{fig:UTOPIA-IGMion:heat_radius-parameters}), increasing for those combinations of parameters that produce a higher fraction of low-to-high energy electrons.

\begin{figure}

\includegraphics[width=\columnwidth]{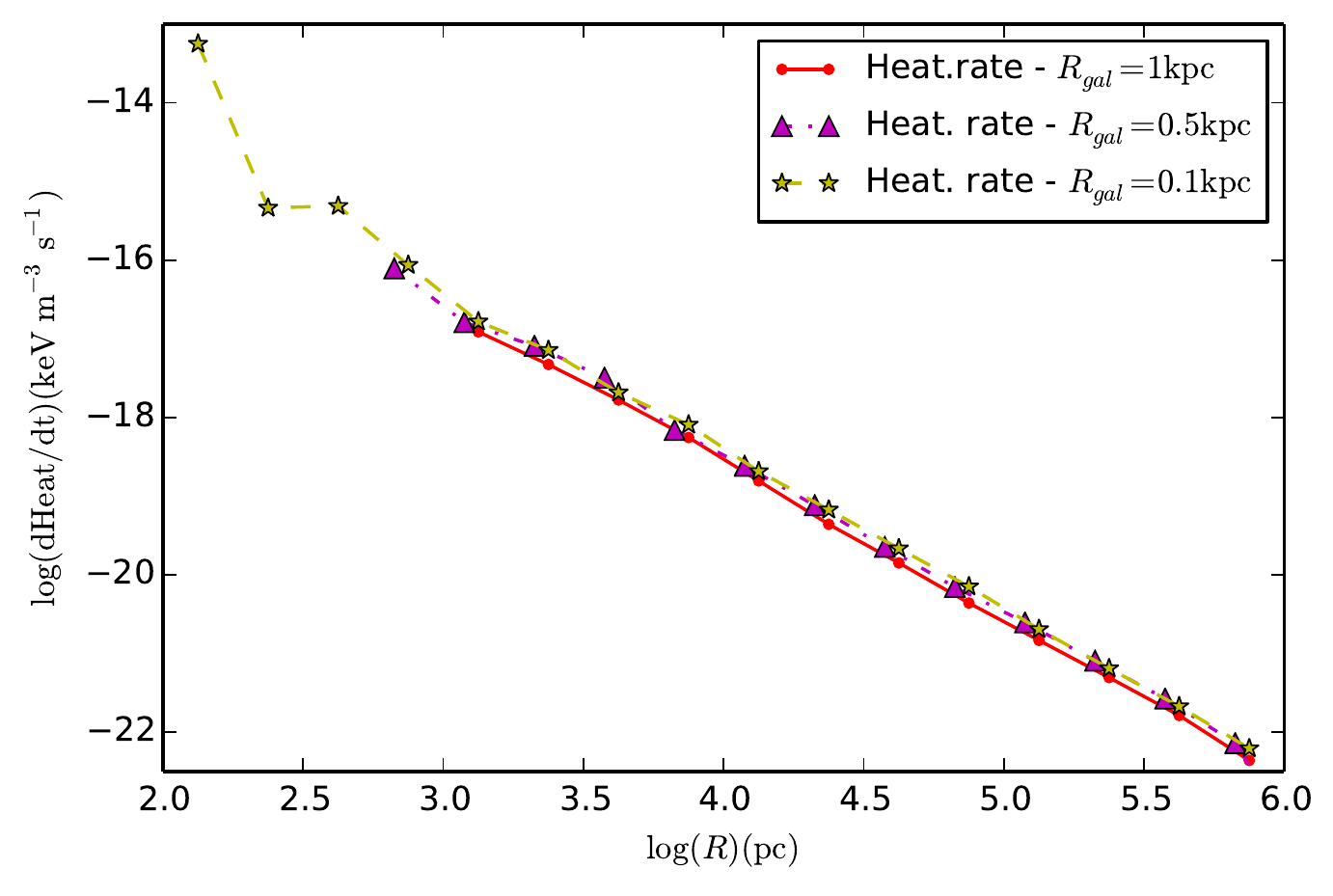}
\caption{Heating rate density as a function of the distance from the centre of  a galaxy at $z = 10$, for different galaxy radii, a fixed $f_{\rm ion} = 0.01$, and our fiducial spectrum of a MQ source with a kinetic luminosity $L_{\rm k} = 10^{40}\,{\rm erg\,s}^{-1}$. Small galaxies are slightly better heating sources, because more low-energy electrons can escape from them.}
\label{fig:UTOPIA-IGMion:heat_radiusgal}
\end{figure}

Fig.~\ref{fig:UTOPIA-IGMion:heat_radiusgal} presents the heating rates per unit volume for different radii of the modelled galaxy. The variation is weak, increasing the heat deposited in the IGM slightly as the galactic radius decreases. This is due to the larger escape fraction for low-energy electrons. The heating rate density decreases with increasing distance from the source, reaching higher values for the smallest galaxy.

\subsection{Total energy deposition}

To complete our picture of reionisation at short scales, it is interesting to analyse the energy budget of the process. For our fiducial MQ model with a luminosity $L = 10^{40}\, {\rm erg\, s}^{-1}$ ($f_{\rm ion} = 0.01$ and $R_g=1 \rm{kpc}$), the typical energy reaching the IGM is 65\% of that emitted by the source (the other 35\% is deposited inside the 1 kpc-galaxy), of which only 7.3\% is transferred into the IGM within a 1 Mpc radius (mainly as ionisation and heating). In other words, the amount of energy deposited in the IGM in the 1 Mpc scale represents only $\sim$ 5\% of the total power emitted by our fiducial source. According to Fig.~\ref{fig:jet-IGMdensity:Elossvsfion}, about a third of this energy is used in ionisation. Roughly 60\% of the energy emitted by the source leaves the $1\, {\rm Mpc}$ volume, and is available for ionising larger IGM regions.  

\begin{figure}
\includegraphics[width=\columnwidth]{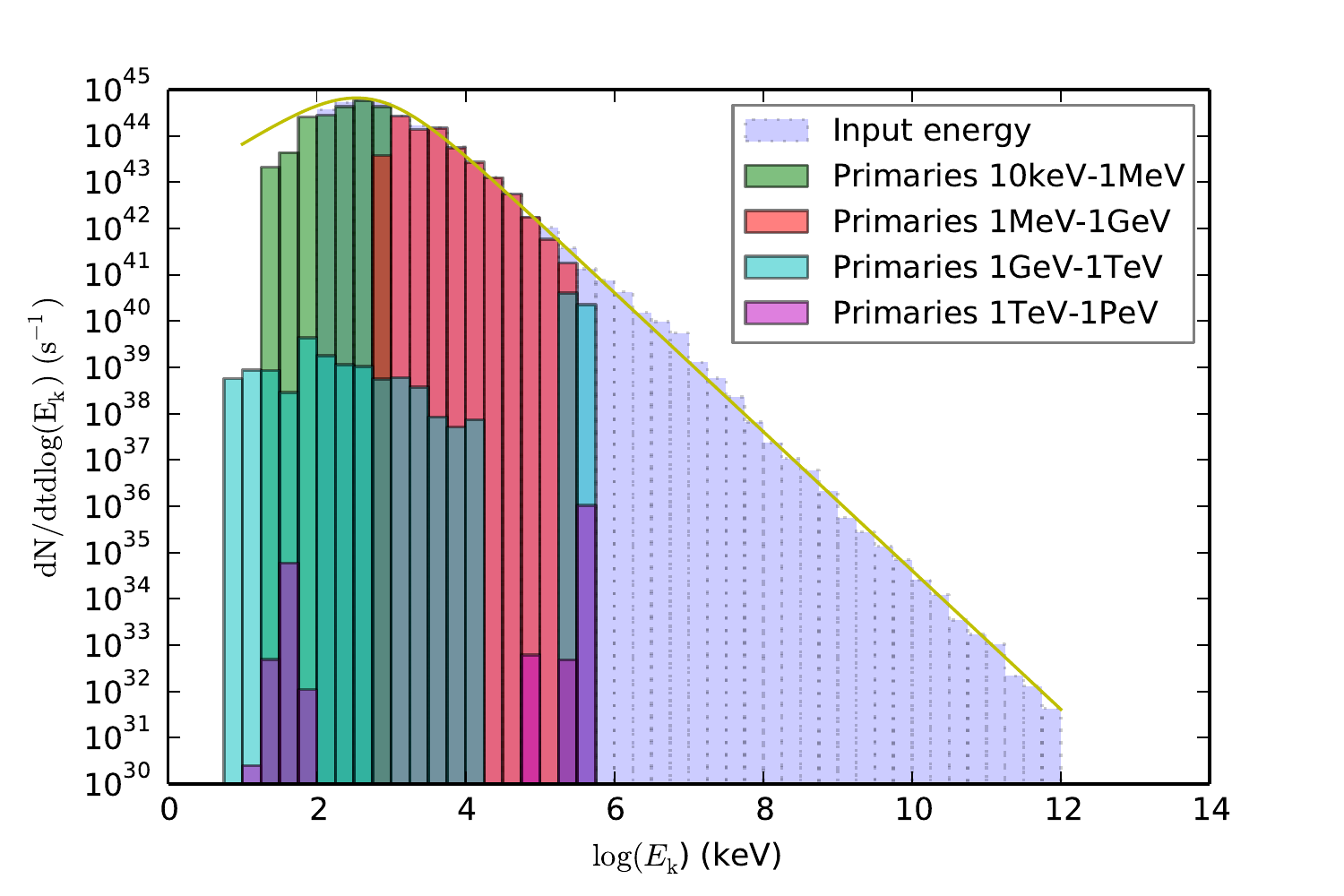}
\caption{Spectral distribution of electrons escaping the simulated $1\,{\rm Mpc}$-radius IGM sphere, injected by a MQ source with a total luminosity of $10^{40}\, {\rm erg\,s}^{-1}$ in a galaxy with $R_{\rm g}=1 \rm{kpc}$ and $f_{\rm ion}=0.01$. Colours represent the kinetic energies of the primary electrons ejected by the MQ. The solid line is the original spectrum of the source. The light-purple (shadowed) histogram is the same as in Fig. \ref{fig:UTOPIA-gxy:histsalidaconancestros}. Ionisation and elastic scattering cool electrons in the low-energy end of the spectrum, whereas IC is responsible for depleting the high-energy end. A new component of low-energy electrons produced by direct Compton scattering enhances the ionisation power of the source.}
\label{fig:outspectrum}
\end{figure}

As it travels through the IGM, the electron spectrum is modified as electrons cool. The major effect is seen in the high-energy end of the spectrum, where IC efficiently eliminates electrons, producing an evolution of the cutoff towards lower energies as particles move away from the source (Fig.~\ref{fig:outspectrum}). Ionisation and elastic scattering continue to work at the lower energy end, but the shape of the spectrum remains similar to the original one. This is because at energies near the peak of the spectrum, particles cool slowly. A new feature not observed at shorter distances, is the presence of a few electrons at keV--MeV energies that come from GeV or TeV. These electrons arise from the direct Compton scattering of high-energy photons produced by IC. A cascade begins to develop, which enlarges the number of low-energy electrons, increasing the ionisation power of the source.

\section{Discussion and conclusions}
\label{conclusions}

Aiming at assessing the contribution of alternative sources to IGM ionisation and heating in the early Universe, we have explored the case of electrons injected by MQ jets \citep[e.g.,][]{heinzsunyaev2002}. We have focused on the effects of these sources on the kpc-Mpc scale, the typical distances between galaxies at redshift 10, and provide quantitative estimations of the transport and deposition of energy by electrons, from the source to the IGM. We modelled a primordial galaxy at $z = 10$, containing MQ sources (one or more) with a total jet kinetic power of $10^{40}\,{\rm erg\, s}^{-1}$, but our results can be easily extrapolated to other astrophysical scenarios regarding the number and luminosity of the sources. Following the guidelines of the pioneering works of \citet{shull1979}, \citet{shull1985}, \citet{valdesferrara2008}, and \citet{furlanettostoever2010}, we have developed self-consistent Monte Carlo simulations of the propagation of escaping electrons through the ISM and IGM spanning 15 orders of magnitude in energy, from $1\,{\rm eV}$ to $1\,{\rm PeV}$. The simulations include only the relevant processes for electron transport, and for collisional ionisation and heating. We compute the energy deposition in the quoted media, paying special attention not only to the energy budget, but also to the spatial distribution of the processes.

Our results can be summarised as follows:
\begin{enumerate}
\item
Low-energy electrons from the sources (up to several tens of keVs), which are those that provide most of the ionisation power, cannot escape primeval galaxies. This agrees with previous works that discuss the corresponding process in our Galaxy \citep[e.g.,][]{Atoyan95}. This behaviour is also consistent to that of low energy X-rays \citep[e.g.][]{xu2014}. Therefore any effective mechanism for IGM reionisation that relies on electrons as ionisation agents, must operate in a two-stage process: energy must be transported through the ISM of galaxies by any sort of carriers that can escape, and then low-energy electrons must be produced locally in the IGM by these carriers.

\item
The escape fraction of electrons from galaxies depends not only on energy, but also on the source spectrum, and may be greater than unity. This is due to cooling processes that decrease the energy of the electrons and produce new ones. The source spectrum is then a key ingredient for understanding reionisation. For a typical MQ, the fraction of the incident energy that escapes from our modelled galaxy grows from null at some tens of keV to almost unity at MeV-GeV energies. Ionisation and elastic scattering within the galaxy are responsible for this increase. The escape fraction falls again to null for TeV energies, due to IC scattering.

\item
Cooling by ionisation and elastic scattering produces a steady transport from intermediate (MeV) to low (keV) energies, both within and outside the galaxies. This process keeps producing low-energy (high ionising/heating power) electrons along the path. The particle population behaves, as it travels, as a large reservoir of energy leaking to the IGM. The energy is stored in high-energy electrons, and the leakage proceeds through low-energy ones. 

\item
High-energy (tens of MeV to $\sim$100 GeV) electrons transport energy far from the source while cooling through IC whereas electrons of a few tens of TeV and up lose their energy fast through IC processes producing a cascade. However, as pointed out by \citet{tueros2014}, this cascade survives the EoR, hence only a fraction of their energy is effectively available for ionisation/heating.

\item
Under the assumed conditions and considering a pre-heated medium, MQs would be able to maintain by themselves large ionisation fractions (of the order of $0.1$) near the galaxies, in the kiloparsec scale. The ionisation rate drops as the square of the distance from the source. The heating rate displays the same behaviour.

\item
In consonance with points i--iv, ionisation and heating rates produced by MQs in the IGM depend mainly on the fraction of low-energy electrons in the jet spectrum. MQs with soft jet spectra are better ionisers and heaters than those with harder spectra. In this sense, the soft component produced from bulk jet electrons proposed by \citet{heinzsunyaev2002}, may play an important role in reionisation.

\item
The total energy used in ionising and heating the IGM at short scales is only a minor fraction of the total energy carried by electrons. This implies that the presence of more effective cooling mechanisms could greatly enhance the ionisation and heating rates obtained in this work.

\end{enumerate}

The aforementioned results are conservative, for three reasons. First, we have neglected the effects of photoionisation. A large amount of low- and high-energy photons are produced mainly by IC, that should add some ionisation to our values. In this sense, our results are a lower bound for the effect of MQs on the IGMs. Second, for time reasons, the fate of photons and electrons leaving the $1\,{\rm Mpc}$ volume has not been investigated. This could produce ionisation at larger distances. Third, we have considered only the basic electron cooling processes, resulting in a low transfer of energy from carriers to ionising electrons. Other processes should add to this transfer, resulting in more ionising and heating power. Among the processes not taken into acccount, IC with EBL photons may be interesting because the higher energy of these photons with respect to CMB ones, implies that the Klein-Nishina regime applies for lower electron energies. Another key effect not taken into account formerly is that of magnetic fields. These are very efficient coolers for electrons, through synchrotron emission. We have not considered other carriers (e.g., $\gamma$ rays from MQs, etc.) either. We will explore the effects of these processes in future works.

The comparison of our results with previous works is not straightforward, because we focus on shorter scales than them \citep{tueros2014,madaufragos2017,leite2017}. If our behaviour could be extrapolated to larger scales, a set of homogeneously distributed galaxies containing MQs (for our fiducial model) would produce mean ionisation rates of the order of $\rm{d}n_{ion}/\rm{d}t\sim 8 \times 10^{-18} \rm{m^{-3} \ s^{-1}}$. This value is of the order of the typical estimates of \citet{leite2017} for CRs from SNe ($\rm{d}n_{ion}/\rm{d}t\sim 2 \times 10^{-18} \rm{m^{-3} \ s^{-1}}$), but much lower than UV photons from Pop II stars \citep[$\rm{d}n_{ion}/\rm{d}t\sim 10^{-14} \rm{m^{-3} \ s^{-1}}$,][]{leite2017}. 

Analogously, the heating rate due to electrons from MQs that we obtain ($\rm{d}Heat/\rm{d}t\sim 8 \times 10^{-20} \rm{keV \ m^{-3} \ s^{-1}}$) is also similar to that of \citet{leite2017} ($\rm{d}Heat/\rm{d}t\sim 1 \times 10^{-19} \rm{keV \ m^{-3} \ s^{-1}}$) for CRs.

Our heating rate corresponds to a global IGM  temperature change of $\Delta T=(2/3)(\rm{d}Heat/\rm{d}t)/(n_H k_B H(z))\sim 50 \rm{K}$, which is of the same order of the increment on the average IGM temperature due to CRs calculated by \citet{leite2017} and similar to that predicted by \citet{madaufragos2017} for X-rays from XRBs, at z$\sim$10. Although our approach is different from theirs and roughly comparable, this implies that the contribution of electrons from MQs to IGM heating could not be neglected.

It should be also pointed out that the increase in the temperature of the medium around the source due to electrons would cause a decrease in the local recombination rate, which would imply a higher ionizing power\footnote{At the same time, the ionisations near the source cause an increase on the number of free electrons which also could contribute to local heating. However this change is negligible in the kpc scale.}. Consequently, the calculated ionisation rates could be underestimated. However, the simulation of this coupling requires a more detailed model, and a large amount of computational time.

The main conclusions derived from our results are the following. Electrons from  MQs would contribute to ionisation significantly only in the kpc scale near the galaxies, whereas their effect on the thermal history of the IGM would rival that of other sources (e.g. X-rays from XRBs and cosmic rays from SNe). This does not mean that they should be neglected regarding ionisation; they could contribute to keep the near medium partially ionised, allowing UV photons to travel farther into the IGM to proceed with reionisation. Among MQs, those with soft spectra display the greatest ionising power.  

Finally, it is important to note that, given the competition between different ionisation processes, the final picture of the EoR can be obtained only by constructing a complete astrophysical scenario that includes all the main sources, and treats all the scales together. We are still far from this picture. Together with the caveats presented in the preceding paragraphs, the main uncertainty source is the lack of knowledge on the properties and evolutionary stages of the first stars and galaxies responsible for reionisation. Both observations with the present and next generation of instruments (e.g., LOFAR, HERA, SKA), together with a detailed physical and astrophysical modelling like the one presented in this paper, are needed to make advances in the comprehension of the main cosmological phase transition that took place during the EoR.

\section*{Acknowledgements}

We are thankful to the anonymous reviewer for the useful suggestions that helped to improve our manuscript. We are deeply indebted to Dr. Susana E. Pedrosa, who provided us with time allocation in the Fenix computational facility at IAFE, and to Dr. Dar\'{\i}o Mitnik for helpful discussions. We wish to acknowledge support from Universit\'e Sorbonne Paris Cit\'e, through the grant ``l'Antenne Argentine de
l'USPC'' (2015), and Argentine CONICET through grant PIP 2014/0265. 

%%%%%%%%%%%%%%%%%%%%%%%%%%%%%%%%%%%%%%%%%%%%%%%%%%



\bibliographystyle{mnras}
\bibliography{Dounaetal2017}

%%%%%%%%%%%%%%%%%%%%%%%%%%%%%%%%%%%%%%%%%%%%%%%%%%

%%%%%%%%%%%%%%%%% APPENDICES %%%%%%%%%%%%%%%%%%%%%

%%%%%%%%%%%%%%%%%%%%%%%%%%%%%%%%%%%%%%%%%%%%%%%%%%

% Don't change these lines
\bsp	% typesetting comment
\label{lastpage}
\end{document}